\documentclass[twocolumn,twocolappendix]{./AAsv6}

\usepackage{graphicx}	
\usepackage{epstopdf}
\usepackage{amsmath}	
\usepackage{amssymb}	
\usepackage{comment}
\usepackage{newtxtext,newtxmath}

\gdef\1054{MS\,1054--03}

\gdef\Ha{H$\alpha$}
\def\farcs{\hbox{$.\!\!^{\prime\prime}$}}
\def\arcs{^{\prime\prime}}
\def\simgeq{{\raise.0ex\hbox{$\mathchar"013E$}\mkern-14mu\lower1.2ex\hbox{$\mathchar"0218$}}} 
\def\sersic{S\'{e}rsic}
\def\H160{$H_{\rm{160}}$}
\def\Re{$R_{\rm{e}}$}

 \usepackage{multirow}
\def\kd{KMOS\textsuperscript{3D}}

\def\kms{km s\textsuperscript{-1}}

\def\rturn{$r_{\rm{turn}}$}
\def\sig{$\sigma$}
\def\sig0{$\sigma_0$}

\def\rc{rotation curve}
\def\rcs{rotation curves}
\def\a3d{ATLAS\textsuperscript{3D}}

\def\vrot{$V_{\rm{max}}$}

\def\rturn{$R_{\rm{turn}}$}
\def\rturnm{$R_{\rm{turn}}^{\rm{morph}}$}
\def\rturno{$R_{\rm{turn}}^{\rm{meas}}$}
\def\ret{$R_{\rm{turn}}^{\rm{morph}}/R_{\rm{e}}$}
\begin {document}

\title {Falling outer rotation curves of star-forming galaxies at $0.6 \lesssim z \lesssim 2.6$ probed with KMOS\textsuperscript{3D} and SINS/zC-SINF}

 \author{Philipp Lang\altaffilmark{1,2}, Natascha M. F{\"o}rster Schreiber\altaffilmark{1}, Reinhard Genzel\altaffilmark{1,3,4}, Stijn Wuyts\altaffilmark{5}, Emily Wisnioski\altaffilmark{1}, Alessandra Beifiori\altaffilmark{1,6}, Sirio Belli\altaffilmark{1}, Ralf Bender\altaffilmark{1,6}, Gabe Brammer\altaffilmark{7}, Andreas Burkert\altaffilmark{1,6}, Jeffrey Chan\altaffilmark{1}, Ric Davies\altaffilmark{1}, Matteo Fossati\altaffilmark{1,6}, Audrey Galametz\altaffilmark{1}, Sandesh K. Kulkarni\altaffilmark{1}, Dieter Lutz\altaffilmark{1}, J. Trevor Mendel\altaffilmark{1}, Ivelina G. Momcheva\altaffilmark{7}, Thorsten Naab\altaffilmark{8}, Erica J. Nelson\altaffilmark{1}, Roberto P. Saglia\altaffilmark{1,6}, Stella Seitz\altaffilmark{6}, Sandro Tacchella\altaffilmark{10}, Linda J. Tacconi\altaffilmark{1}, Ken-ichi Tadaki\altaffilmark{1}, Hannah \"{U}bler\altaffilmark{1}, Pieter G. van Dokkum\altaffilmark{9}, 
 David J. Wilman\altaffilmark{1,6}}

 \altaffiltext{1}{Max-Planck-Institut f\"{u}r extraterrestrische Physik, Giessenbachstrasse, D-85748 Garching, Germany}
 \altaffiltext{2}{Max-Planck-Institut f\"{u}r Astronomie, K\"{o}nigstuhl 17, D-69117 Heidelberg, Germany}
 \altaffiltext{3}{Department of Physics, Le Conte Hall, University of California, Berkeley, CA 94720, USA}
 \altaffiltext{4}{Department of Astronomy, New Campbell Hall, University of California, Berkeley, CA 94720, USA}
 \altaffiltext{5}{Department of Physics, University of Bath, Claverton Down, Bath, BA2 7AY, UK}
 \altaffiltext{6}{Universit\"{a}ts-Sternwarte M\"{u}nchen, Scheinerstr. 1, M\"{u}nchen, D-81679, Germany}
 \altaffiltext{7}{Space Telescope Science Institute, 3700 San Martin Drive, Baltimore, MD 21218, USA}
 \altaffiltext{8}{Max-Planck-Institut f\"{u}r Astrophysik, Karl Schwarzschildstr. 1, D-85748 Garching, Germany}
 \altaffiltext{9}{Astronomy Department, Yale University, New Haven, CT 06511, USA}
\altaffiltext{10}{Institute of Astronomy, Department of Physics, Eidgen\"{o}ssische Technische Hochschule, ETH Z\"{u}rich, CH-8093, Switzerland}

\begin{abstract}


We exploit the deep resolved \Ha\ kinematic data from the \kd and SINS/zC-SINF surveys to examine the largely unexplored outer disk kinematics of star-forming galaxies {(SFGs)} {out to the peak} of cosmic star formation.  Our sample contains 101 {SFGs} representative of the {more massive} ($9.3 \lesssim \log{M_*/M_{\sun}} \lesssim 11.5$) main sequence population at $0.6 \leqslant z \leqslant 2.6$.  
{Through a novel stacking approach we are able to constrain a representative rotation curve extending out to $\sim 4$ effective radii.  This average rotation curve exhibits a significant drop in rotation velocity beyond the turnover, with a slope of $\Delta V / \Delta R = -0.26^{+0.10}_{-0.09}$ in units of normalized coordinates $V/V_{\rm{max}}$ and $R/R_{\rm{turn}}$.  This result confirms that the fall-off seen previously in some individual galaxies is a common feature of our sample of high-z disks.}  We show that this outer fall-off strikingly deviates from the flat or mildly rising rotation curves of local spiral galaxies of similar masses.
We furthermore compare our data with models including baryons and dark matter demonstrating that the falling stacked rotation curve can be explained by a high mass fraction of baryons relative to the total dark matter halo ($m_{\rm{d}} \gtrsim 0.05$) in combination with a {sizeable} level of pressure support in the outer disk.  These findings are in agreement with recent studies demonstrating that star-forming disks at high redshift are strongly baryon dominated within the disk scale, and furthermore suggest that pressure gradients caused by {large turbulent gas motions are present} even in their outer disks.   We demonstrate that these results are largely independent of our model assumptions such as {the} presence of a central stellar bulge, the effect of adiabatic contraction at fixed $m_{\rm{d}}$, and variations in the concentration parameter.

\end{abstract}

\section {Introduction}
\label{intro.sec}

Understanding the relative distribution of baryons and dark matter (DM) in galaxies is a key ingredient towards validating and constraining current galaxy evolution models.
Observationally, disentangling the contribution of luminous and DM at different radii has been most successful through analyzing the shape of observed rotation curves (RCs), also used in conjunction with information from gravitational lensing (see e.g. Courteau et al. 2014).  
Since the discovery of {large rotation velocities} in the outer disks of nearby spirals {(e.g. Babcock 1939)}, it has been confirmed that rotation curves of disk galaxies are flat in their outer parts beyond the peak reached at $R_{\rm{max}}$ (e.g. Sofue \& Rubin 2001 and references therein), leading to a global picture in which disk galaxies are baryon dominated in their inner parts and DM dominated in their outer regions (Courteau \& Dutton 2015, and references therein). {Efforts {to quantify} the dark matter contribution of galaxies on the scale of their optical disks have revealed that the baryonic fractions at fixed radius increase with stellar mass.}  While 
local low-mass late-type galaxies appear to be dominated by dark matter, massive local spirals like  the Milky Way are baryon-dominated within their optical disks (Martinsson et al. 2013; Bovy \& Rix 2013; Barnab\`{e} et al. 2011, 2012; Courteau \& Dutton 2015). 

{Spatially resolved kinematics of galaxies at the peak epoch of cosmic star formation density, {$z\sim 1 - 2$}, is now feasible with} the advent of new integral-field-unit (IFU) spectrographs operating in the near infrared (NIR) in the last decade.  {These} instruments enabled surveys of ionized gas kinematics, such as SINS/zC-SINF (F{\"o}rster Schreiber et al. 2009; Mancini et al. 2011), MASSIV (e.g. \'Epinat et al. 2009, 2012), AMAZE/LSD (Gnerucci et al. 2011), and {SHiZELS} (e.g. Swinbank et al. 2012) with SINFONI on the Very Large Telescope (VLT).  More recently, the multiplexing capabilities of the 24-IFU KMOS instrument at the VLT (Sharples et al. 2012) now allows one to efficiently map the two-dimensional kinematics of much larger and more complete samples.  
Surveys such as \kd (Wisnioski et al. 2015) and KROSS (Stott et al. 2014, 2016) are systematically uncovering the dynamical properties of $> 1000$ SFGs at $z \sim 1 -3$.  The surveys robustly confirm earlier findings that a majority of SFGs, which lie on a tight  `main sequence' (MS) between their stellar mass and star formation rate (SFR; e.g. Noeske et al. 2007; Elbaz et al. 2007; Daddi et al. 2007; Whitaker et al. 2012, 2014) {have kinematics dominated by rotation} with regular disk-like velocity fields.  Their disk velocity dispersions  {are large} and increase with redshift, reaching up to $\sim 60-70$ \kms at $z\sim2$,  {as expected} for marginally stable gas-rich disks (see e.g. F{\"o}rster Schreiber et al. 2006, 2009; Genzel et al. 2006, 2008; Law et al. 2009; \'Epinat et al. 2009, 2012; Kassin et al. 2012; Wisnioski et al. 2015; Tacconi et al. 2010, 2013; Daddi et al. 2010; Genzel 
et al. 2015).

SFGs on the main sequence are moreover found to have exponential light and mass profiles (e.g. Wuyts et al. 2011b; Bell et al. 2012; Lang et al. 2014), supporting an underlying smooth stellar disk structure.  {Additionally, bulge-disk decompositions show that the most massive ({$\log{M_*/M_{\sun}} > 11$}) SFGs at high redshift have significant central bulges with bulge-to-total ($B/T$) ratios up to $40 - 50 \%$ (Lang et al. 2014; see also, e.g., Bruce et al. 2014, Tacchella et al. 2015a,b).}

So far, kinematic probes of SFGs at high redshift based on IFU  {and} long-slit observations are mostly limited {to the regions within} the peak of their rotation curves due to the fast decline of \Ha\ surface brightness with increasing galactocentric radius.  
Comparing the stellar plus inferred gas mass relative to the dynamical mass inside their optical disks, studies found that {SFGs} at larger look-back times are more baryon-dominated, with the baryon-to-dark matter fraction at a given stellar mass increasing with redshift (F{\"o}rster Schreiber et al. 2009; van Dokkum et al. 2015; Burkert et al. 2016; Wuyts et al. 2016; Price et al. 2016, Stott et al. 2016).  Extrapolating the estimate on the inner dark matter and baryonic distributions employing the {approach of Mo et al. (1998)} based on a large compilation of IFU datasets, Burkert et al. (2016) determined that the fraction of the baryonic mass in the disk compared to the total amount of dark matter within the virial radius ($m_{\rm{d}} = M_{\rm{baryonic}}/M_{\rm{Halo}} (< r_{200})$) is $\sim 0.05$ for massive high-z SFGs at  $0.5 < z < 2.5$. 

These high baryonic fractions might to first order reflect more compact baryonic configurations {(in line with the smaller size at fixed mass for the stellar component of high-z star-forming disks; e.g. van der Wel et al. 2014a)} as well as different underlying dark matter halo structure at higher redshifts compared to local galaxies.  The latter is suggested by cosmological dark matter simulations, which predict a decrease of halo concentration with increasing redshift (e.g. Neto et al. 2007; Zhao et al. 2009; Prada et al. 2012).
Although applied to statistically firm samples,  {robust determinations of baryonic fractions at high redshift are limited by uncertainties in both stellar and gas mass measurements.}  Moreover, it is still unclear how the infall of baryons acts on the structure of dark matter halos, such as proposed by the adiabatic contraction scenario (e.g. Blumenthal 1986; Gnedin et al. 2004 and references therein).  While the impact of adiabatic contraction on local rotation curves  {remains uncertain} (e.g. Dutton et al. 2005; Kassin et al. 2006; Schulz et al. 2010; Auger et al. 2010; Puglielli et al. 2011), the effect of rapid outflowing gas due to supernova or AGN feedback might even lead to a reduction of contraction or act to expand dark matter halos (Navarro et al. 1996; Gnedin \& Zaho 2002; Read \& Gillmore 2005).

An independent method to study the dark matter fraction in galaxies is abundance matching, where observational constrains on the stellar mass function at different redshifts are used in conjunction with results of cosmological dark matter simulations to match the cumulative abundance of galaxies to that of dark matter haloes. Assuming a universal IMF, the inferred {fraction} of stellar mass in the disk compared to the mass of the entire halo {is} dependent on halo mass, with a peak of $m_{\rm{d}} \sim 0.02 - 0.025$ (Moster et al. 2013; Behroozi et al. 2010, 2013) at $1 < z < 3$.  Accounting for the {higher} gas fractions at high redshift, this implies $m_{\rm{d}} \sim 0.05$ (see discussion by Burkert et al. 2016). 

The high baryonic fractions of disk galaxies at high redshift likely leave an imprint on their rotation curves.  {This has been confirmed on the basis of six individual galaxies with very deep integrations, exhibiting a {drop in their rotation} curve beyond $R_{\rm{max}}$ {(Genzel et al. 2017)}.  In this paper, we follow-up on this result based on a larger sample, to explore how common dropping outer disk rotation curves are among high-z SFGs.}  Due to the sharply dropping \Ha\ surface brightness with galactocentric radius, we approach this by stacking the {extensive set of IFU data from the deep} \kd and SINS/zC-SINF surveys.  The combination of these datasets provides a unique sample of 101 galaxies with a good coverage of the SFR - stellar mass plane at $0.6 \leqslant z \leqslant 2.6$.  
{Based on our stacked \rc, we discuss implications on dark matter fractions of high-z outer disk regions, which are largely independent of gas mass {estimate} and assumptions on the IMF.}

The paper is structured as follows.  In Section 2, we give an overview of the observations and sample.  The extraction of spectra, their normalization, and the final stacking, is described in Section 3. We present our stacking results in Section 4, {together with a comparison with dark matter halo models. We discuss the implications of our findings in Section 5 and summarize our results in Section 6}. 
Throughout this paper, we assume a Chabrier (2003) initial mass function (IMF), and adopt the cosmological parameters $(\Omega _M, \Omega _{\Lambda}, h) = (0.3, 0.7, 0.7)$. 

\section{Observations and sample}
\label{data.sec}

The basis of this analysis is spatially resolved IFU observations of \Ha\ of massive SFGs at $0.6 \leqslant z \leqslant 2.6$ from the \kd survey (Wisnioski et al. 2015) and the SINS/zC-SINF survey (F{\"o}rster Schreiber et al. 2009; Mancini et al. 2011; F{\"o}rster Schreiber et al. in prep.).  Here, we summarize the main properties of those datasets, the ancillary data used, as well as further selection criteria and the properties of our sample.

\subsection{The \kd and SINS/zC-SINF datasets}

The \kd survey is an ongoing program to observe galaxies at $0.6 < z < 2.7$ using the multiplexed IFU instrument KMOS on the VLT (Sharples et al. 2012).  The targets are drawn from the 3D-HST Treasury Survey with the Hubble Space Telescope (HST; Brammer et al. 2012; Skelton et al. 2014; Momcheva et al. 2016) in the CANDELS HST imaging survey fields (Grogin et al. 2011; Koekemoer et al. 2011) accessible from the VLT: GOODS-South, COSMOS, and UDS.  The selection is primarily based on stellar mass ($\log(M_{\star}) > 9.5$).  Additional  {selection criteria are }a magnitude cut ($K_{s,AB} < 23$ mag), the availability of a reliable and sufficiently accurate redshift (from spectroscopy or HST grism data), and the avoidance of sky line contamination or low atmospheric transmission around the wavelength of the {redshifted} \Ha\ emission.  The observations are done using either the $YJ$, $H$ or $K$ band grating to target galaxies at $ z \sim 0.9$, $z \sim 1.5$, and $z \sim 2.3$, respectively.  The survey strategy 
emphasizes deep, high $S/N$ observations of individual galaxies; the median integration times are $\sim 4.7$ hrs, $\sim 7.8$ hrs, and $\sim 8.3$ hrs for galaxies observed in the $YJ$, $H$, and $K$ band, respectively.  Galaxies with fainter line fluxes, such as high-mass objects well below the MS have longer integrations (up to $\sim25$ hrs).  Each KMOS IFU has a field of view (FOV) of $2\farcs8$ x $2\farcs8$ in size, sampled by $0\farcs2$ per spaxel.  Typical seeing conditions yield values for the full width half maximum (FWHM) resolution of $0\farcs4$ - $0\farcs6$ {(corresponding to {$\sim 2.7 - 5$} kpc within the redshift range considered here)}.  The spectral resolution depends on the observing band and IFU, and lies between $\sim 60$ and $\sim 110$ \kms\ FWHM.  {For this analysis we use data from \kd observed up to April 2016, including $\sim 564$ observed targets, of which 425 are detected in \Ha, and 359 are furthermore spatially resolved}\footnote{{Here, and throughout this paper, we define spatially resolved galaxies as targets where the detected \Ha\ emission extends beyond one resolution element.}}.\\

The SINS/zC-SINF survey (F{\"o}rster Schreiber et al. 2009; Mancini et al. 2011; F{\"o}rster Schreiber et al. in prep.) is a program using the IFU instrument SINFONI (Eisenhauer et al. 2003; Bonnet et al. 2004) at the VLT, targeting $\sim 110$ {SFGs} at $z \sim 2$.  For our study we use a subset of 35 targets observed in adaptive optics (AO)-assisted mode, which we refer to as the SINS/zC-SINF AO sample below.  For several targets, these data  {are} supplemented by deep observations with SINFONI in seeing-limited mode.  The target selection for SINS/zC-SINF is based on optical spectroscopic surveys with a variety of primary magnitude and/or color cuts, yielding a sample {that probes well} the $z \sim 2$ main sequence population of massive SFGs (see F{\"o}rster Schreiber et al. 2009 and Mancini et al. 2011 for details).  
The FWHM of the spatial PSF is $\sim 0\farcs6$ {({$\cong 5$} kpc)} in natural seeing mode, and ranges between $0\farcs15$ and $0\farcs25$ {({$\cong 1.2 - 2.1$} kpc)} in AO mode, with a spectral resolution between $\sim 80$ and $ \sim 120$ \kms FWHM.  The spatial sampling is $0\farcs125$/spaxel for the seeing-limited data and $0\farcs05$/spaxel for the AO-assisted data.  Depending on the observing strategy followed for the sky and background subtraction, the effective FOV of the reduced data ranges between $\sim$ $2\arcs$ x $2\arcs$ and $\sim$ $4\arcs$ x $4\arcs$.

\subsection{Ancillary data}
\label{ancillary.sec}
In addition to the IFU datasets, we make use of the available wealth of ancillary data in the COSMOS, GOODS-S and UDS fields providing information on integrated and resolved properties of our galaxy sample.\\
 {We derive stellar masses and SFRs following the prescriptions in Wuyts et al. (2011a)}.  Model assumptions include the Chabrier (2003) IMF, a solar metallicity, the Calzetti et al. (2000) reddening law, and constant or exponentially declining star formation histories.  The SEDs were constructed from optical, near-IR, and mid-IR $3-8 \mu m$ photometry available in all relevant fields (for details and references, see Skelton et al. 2014 for the 3D-HST catalogs used for the \kd sample, and F{\"o}rster Schreiber et al. 2009 and Mancini et al. 2011 for the photometry used for the SINS/zC-SINF sample), supplemented by mid- and far-infrared photometry from $Spitzer$/MIPS and $Herschel$/PACS (Lutz et al. 2011; Magnelli et al. 2013; Whitaker et al. 2014).  {Molecular gas masses for our sample are estimated using the scaling relations from Genzel et al. (2015) and updated in {Tacconi et al. (2017)}}\footnote{{Using instead the gas mass scaling relations from Genzel et al. (2015) yield average gas mass estimates for our sample that are higher by $\sim 0.07 \log{M_{\rm{gas}}/M_{\sun}}$.}}.\\

We also use the available high-resolution panchromatic HST imaging from the CANDELS survey (Grogin et al. 2011, Koekemoer et al. 2011), and the morphological parameters derived from the $H$-band \sersic\ profile fitting from van der Wel et al. (2012), supplemented by available $H$-band HST WFC3 and NICMOS imaging for our SINS/zC-SINF AO targets (Tacchella et al. 2015b; see also F{\"o}rster Schreiber et al. 2011).\\
 {We adopt the morphological parameters from $H$-band images (i.e. covering the rest-frame NIR to optical regime), since these are the best proxies} for the baryonic distributions for our galaxy sample.  {While the stellar mass distribution} of high-z SFGs is shown to be more concentrated than the $H$-band light,  it is important to quantify the structure of the \textit{baryonic} distribution (i.e. stellar and gas) for our sample.  
 {\Ha\ emission studies of massive SFGs} have shown that the distribution of star formation and thus the inferred distribution of molecular gas (by inverting the Kennicutt-Schmidt law, Kenicutt 1998) is less centrally concentrated than the stellar component and in certain cases exhibits `ring-like' morphologies for the most massive systems (Wuyts et al. 2013; Genzel et al. 2014; Nelson et al. 2012, 2016).  
Therefore, given the significant gas fractions of SFGs at $z \sim 1 - 2.5$, the baryonic mass is expected to be more spatially extended than the stellar mass distribution.  Since we are currently not able to derive robust extinction-corrected SFR and gas distributions for our sample {(see Tacchella et al. 2017 for a discussion on resolved extinction-corrected SFRs for a subset of the SINS/zC-SINF sample), we use the $H$-band morphology as a best proxy for the baryonic distributions for our galaxy sample.}


\subsection{Sample selection and properties}
 \label{stacking_sample.sec}

For our sample selection as well as for the comparison of sample properties presented below, we consider targets from both \kd and the AO-assisted targets from SINS/zC-SINF which are spatially resolved (in the remainder referred to as the `\kd  + SINS/zC-SINF AO samples`), and furthermore specifically select targets that are suitable for our science goals.  The basis of this selection is derived from resolved velocity fields and rotation curves extracted from the data cubes as explained in detail in Section\ \ref{normalization.sec}.\\

The selection criteria are the following:\\

\begin{enumerate}
 \item  {First, we select rotationally supported disk galaxies according to criteria 1 and 2 of Wisnioski et al. (2015).}  Those criteria require galaxies to exhibit an overall smooth and disk-like velocity field with a continuous velocity gradient along the kinematic axis, {and to have {$V_{\rm{rot}}/\sigma_0 > 1$}}{; we define {$V_{\rm{rot}}/\sigma_0$} as the ratio between the intrinsic inclination-corrected peak rotation velocity, $V_{\rm{rot}}$, and the intrinsic velocity dispersion $\sigma_0$)}\footnote{Measurements of $V_{\rm{rot}}/\sigma_0$ for all individual galaxies are extracted from inclination-corrected rotation curves and dispersion profiles {at the outermost radius at which they can be determined. These measurements are furthermore corrected for the remaining level of beam smearing based on beam smearing correction factors that take into account the radius of measurement (see a detailed description in Appendix A in Burkert et al. (2016)).}}.  We note that the vast majority of the final selected galaxies ($> 90\%$) exhibit smooth and centrally peaked dispersion profiles as expected for intrinsic 
rotating disks given the level of spatial beam smearing.  The remaining targets show asymmetries in their dispersion profiles  {with smooth} underlying velocity gradients. Galaxies with signatures of major mergers, or potentially perturbed by a neighbour based on projected distance, redshift, and stellar mass ratio, are rejected.

\item Next, we require the individual extracted rotation curve of a galaxy to exhibit a continuous change of slope with a significant flattening in rotation velocity towards larger radii.  The amount of required flattening is chosen such that the expected maximum velocity $V_{\rm{max}}$ and the corresponding radius \rturn\ can robustly be determined and {does not extend beyond the radius out to which the individual RC can be reliably measured {($R_{\rm{obs}}$), allowing for small extrapolations of $\leqslant 0.2\arcs$ (i.e. corresponding to one spatial pixel in the datacubes)}}. 
By applying this cut, we remove galaxies from our sample either with too limited $S/N$ in their outer parts towards \rturn\ or targets  {that} are too extended such that the flattening of the rotation curve falls outside the FOV. 
\item We furthermore discard galaxies with strong residual contamination of OH sky lines and the atmospheric $\rm{O}_2$ emission feature within the wavelength region of \Ha. 

 \end{enumerate}

\begin {figure*}[htp]
\centering
 \includegraphics[width=0.70\textwidth]{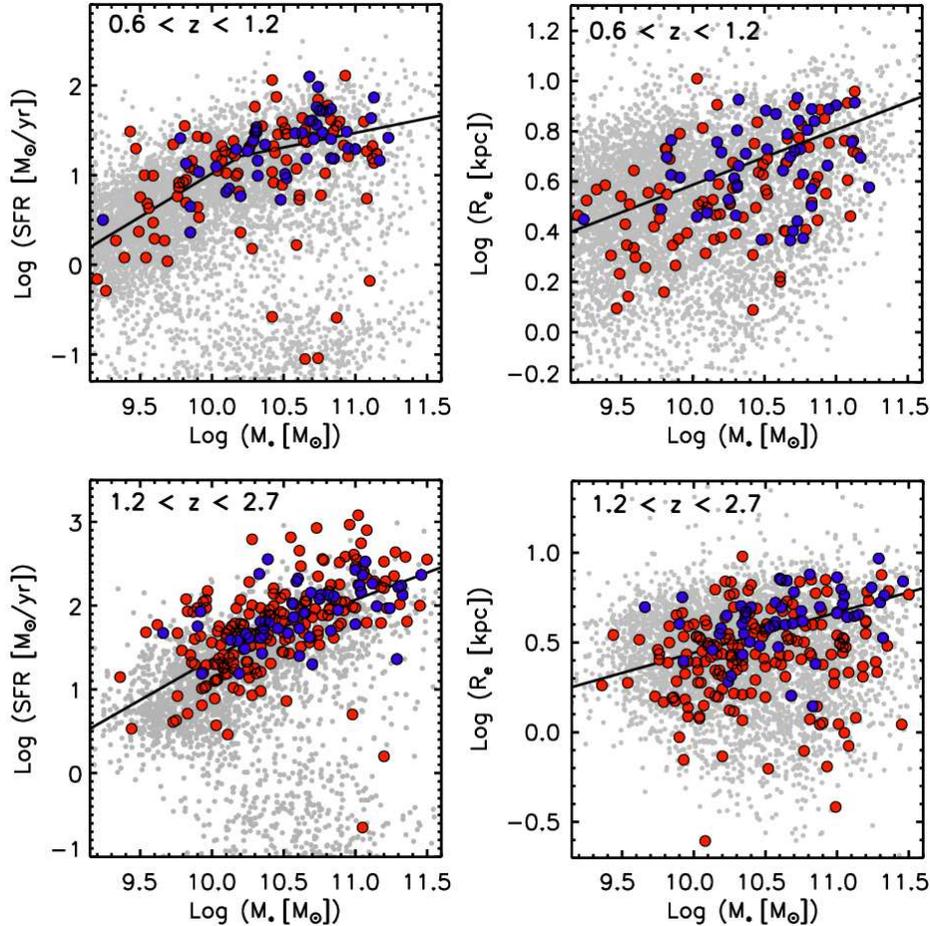}
\caption[Stacked sample in the $R_{\rm{e}} - M_*$, $SFR - M_*$ planes]{Properties of our sample of stacked galaxies (blue symbols) compared to all detected and spatially resolved \kd and SINS/zC-SINF AO galaxies (combined blue and red symbols) , plotted in the $SFR - M_*$  plane (left), and in the $R_{\rm{e}} - M_*$ plane (right).  Effective radii are measured on $H$-band light distributions.  The grey points show the 3D-HST background population in the respective redshift ranges, rejecting targets with purely photometric redshifts.  The solid lines in the left panels represent the broken-power law parametrization of the main sequence by Whitaker et al. (2014) for the corresponding redshifts.  The solid lines in the right panels represent the size-mass relation for late-type galaxies as derived by van der Wel et al. (2014a).}
\label{sample_properties1.fig}
 \vspace{3mm}
\end {figure*}

\begin{table}[t]
\caption[a]{Median properties of the stacking sample.}
{\small
\centering
\begin{tabular}{l*{2}{c}r}
Property & Median &  \\
\hline
 $z$ 				& 1.52 \\
$\log(M_*$[$M_{\sun}$]$)$ 	& 10.61 \\
$\log(M_{\rm{baryonic}} = M_* + M_{\rm{gas}}^a)$[$M_{\sun}$]$)$ 	& 10.87 \\
$SFR$ [$M_{\sun}/yr$] 		& 41.9 \\
$sSFR$ [$Gyr^{-1}$] 		& 1.20 \\
$R_{e}^b$ [kpc] 			& 4.6 \\
$n^b$ 			& 1.1 \\
\hline
\end{tabular}
\tablenotetext{a}{Gas mass estimate based on empirical scaling relations.}
\tablenotetext{b}{Intrinsic effective radius {along the major axis} and \sersic\ index derived from $H$-band \sersic\ fits. }
\label{Stack_prop.tbl}}
\end{table}

After applying Criterion 1, {284} galaxies remain.  We note that the fraction of galaxies with signs of major mergers is relatively small ($\sim 3\%$ of the total sample).  
After applying Criterion 1 and 2, there are 140 galaxies left, implying that Criterion 2 removes the largest portion of the total sample among our selection criteria.  We dedicate a detailed analysis on potential effects connected with our selection that may bias our selected sample by using simulated rotation curves, presented in Appendix C.  In short, we find that the rejection of galaxies according to Criterion 2 is mainly driven by the variations of $S/N$ among the observed data and does not yield significant biases in our stacking results. Finally, with all criteria applied, the final sample (henceforth referred to as the `stacking sample') consists of 101 targets (92 from \kd and 9 from the SINS/zC-SINF AO sample) to be used in our stacking analysis.  48 and 53 of those fall within the redshift ranges $0.6 < z < 1.2$ and  $1.2 < z < 2.6$, respectively.\\

Table 1 summarizes the median values of the main properties of the selected stacking sample.  The quoted baryonic mass $M_{\rm{baryonic}}$ is computed as the sum of the stellar mass and the gas mass estimate.  {Effective radii quoted here and throughout the paper are measured along the major axis.}  Examining the distribution of \sersic\ indices $n$ for the stacking sample, we find a very good agreement with exponential mass profiles.\\
Figure\ \ref{sample_properties1.fig} highlights the key properties of our selected stacking sample in the $SFR - M_*$ and  $R_{\rm{e}} - M_*$ planes for two redshift bins.  Shown are the stacking sample (blue symbols) compared to the \kd + SINS/zC-SINF AO samples of detected and resolved galaxies (combined blue and red symbols), together with the underlying 3D-HST parent sample (grey points).  { Also shown are the stellar mass - SFR relation and the stellar mass - radius relation for late-type galaxies as parametrized by Whitaker et al. (2014) and van der Wel et al. (2014a), respectively, and computed for the average redshift of our sample in each redshift bin considered ({$0.6<z<1.2$ with $<z> = 0.9$, and $1.2<z<2.6$ with $<z> = 2.1$}).}  The stacking sample yields a fair and robust representation of the underlying population of massive main sequence SFGs, judged by its good overlap with the \kd + SINS/zC-SINF AO samples, as well as by a fairly homogeneous coverage of the star-forming main sequence and the size-mass relation for late-type galaxies within the mass range sampled.\\

To quantify our sample properties (including stellar mass, SFR, size, main sequence offset $\Delta(\rm{MS}$), and \sersic\ index) and possible selection biases more thoroughly, we present a detailed statistical analysis in Appendix A.  In short, we find little to no bias in the average examined galaxy properties between the stacking sample and \kd  + SINS/zC-SINF AO samples.  However, the size distribution of galaxies in the stacking sample is offset towards larger values, most significantly for galaxies at $z > 1.2$.  On the basis of our mock analysis presented in Appendix C, we find a qualitatively consistent trend between selection fraction and $R_{\rm{e}}$ which plausibly serves as an explanation for our selection bias towards larger galaxies and in turn stems from a positive correlation between $R_{\rm{obs}}$ and $R_{\rm{e}}$ found in our \kd  + SINS/zC-SINF AO dataset.  The reason for the latter correlation might be that smaller galaxies are affected more severely by beam smearing, 
which hampers a robust measurement of their rotation curve towards the flattening radius and in turn results in removing them from the sample after applying Criterion 2.\\

{
To demonstrate that our sample selection is mostly driven by galaxy size as well as $S/N$, we examine the relation between $R_{\rm{e}}$, $R_{\rm{obs}}$, and the position of the expected turnover in Appendix B. In short, we find that our stacking sample contains targets with overall larger  $R_{\rm{obs}}$ for a given galaxy size (and thus higher $S/N$) than the subset of disks in the \kd  + SINS/zC-SINF AO parent sample.  Moreover, the majority of galaxies in the stacking sample have rotation curves reaching out to radii at or beyond the expected turnover, whereas targets rejected according to Criterion 2 do not reach the expected turnover on average. There is, however, a minor subset of rotation curves that extend beyond the expected turnover radius but without sufficient flattening.  As these cases might represent galaxies with potentially outer rising rotation curves, we include them in our stacking methodology and show that we obtain consistent results compared to when using our fiducial stacking sample of 101 galaxies (see discussion in Appendix B.}\\

\section{Methodology}
\label{method.sec}

\subsection{Extraction and normalization of RCs}
\label{normalization.sec}

\begin {figure}[t]
\centering
 \includegraphics[width=0.49\textwidth]{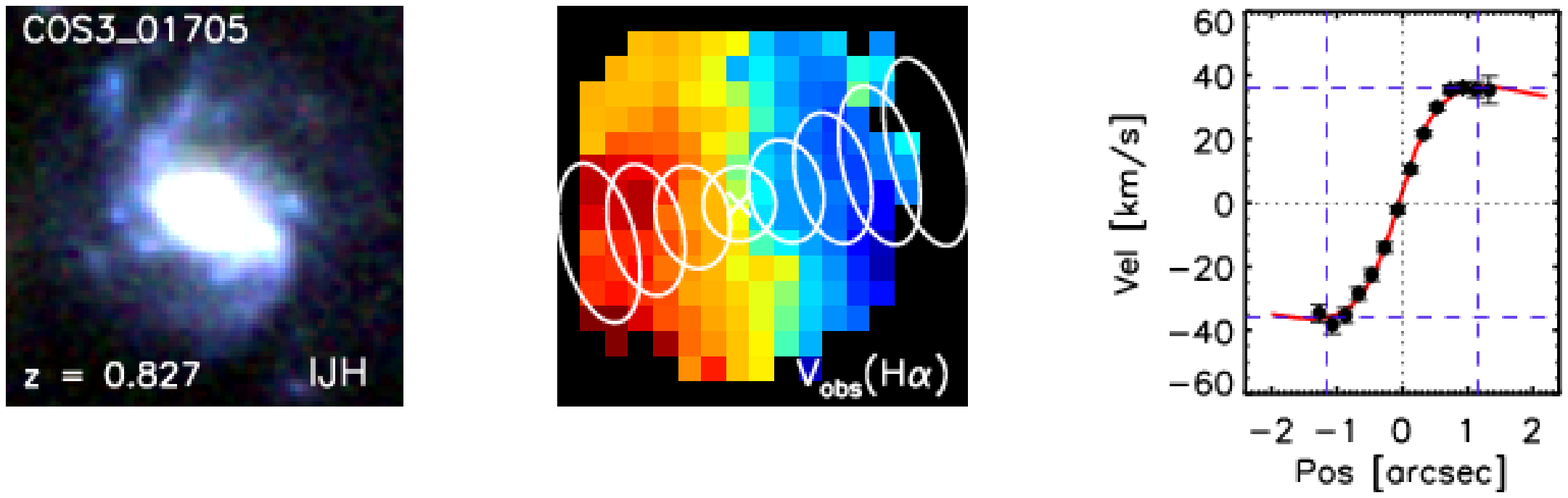}
\includegraphics[width=0.49\textwidth]{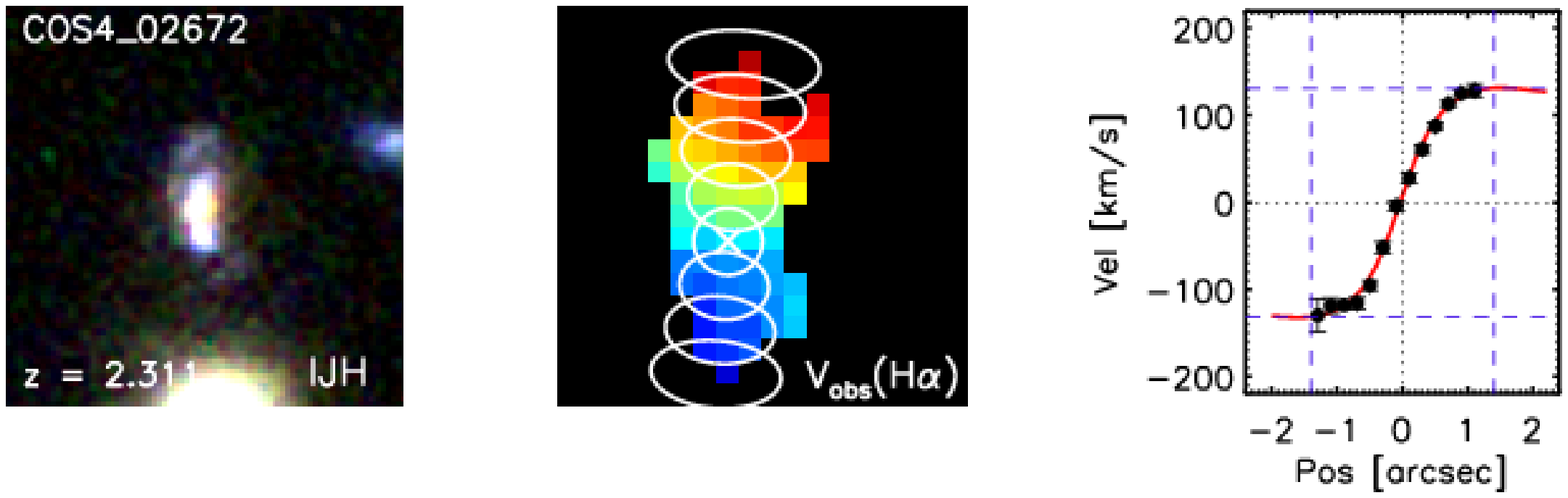}
\caption[Spectral extraction and the normalization of rotation curves]{Examples of the extraction of spectra and the normalization of rotation curves for two targets with different redshifts.  Left panels: $IJH$ color composite from CANDELS/HST imaging.  Middle: \Ha\ velocity field with overplotted apertures used for spectral extraction along the kinematic major axis.  Right: Extracted rotation curve shown as black data points, together with the fitted exponential disk models shown in red.}
\label{apertures.fig}
 \vspace{3mm}
\end {figure}

As a first step, we generate \rcs\ for each galaxy in our sample as follows.  We extract spectra from our data cubes within apertures placed along the kinematic major axis determined from the \Ha\ velocity field over the pixels with sufficient $S/N$ ($\gtrsim 5$).  {The kinematic axis is computed to lie along the largest observed velocity  {gradient} within the velocity field, with the kinematic center being at the midpoint between the velocity minimum and maximum.}  For details on how the kinematic maps are computed, see F{\"o}rster Schreiber et al. (2009), Mancini et al. (2011), and Wisnioski et al. (2015).  To optimize the signal in the spectral extraction along the kinematic axis, we use the following method.  We sum the spectra of individual pixels over apertures with a fixed radial diameter corresponding to the spatial PSF FWHM, such that the velocity field in the radial direction is not additionally smoothed at a significant level.  The size of the apertures in the perpendicular direction is increasing radially towards the fainter regions such that the signal in a given aperture is optimized without additionally smoothing the velocity profile.  
The rate at which the size of the elliptical apertures increases with radius is dependent 
on the inclination angle of the galaxy and gauged on the basis of artificial velocity maps of mock galaxies using rotating exponential disk models.  Figure\ \ref{apertures.fig}  {highlights} the extraction of spectra for two galaxies at different redshifts.  The left panels display the observed-frame $IJH$ color composite to show the rest-frame optical morphology.  In the middle panels, the apertures used for extraction are plotted over the velocity fields.

In order to generate the \rcs\ for each galaxy, we determine the observed \Ha\ velocity as a function of galactocentric radius.  The velocity is measured from fitting a single Gaussian profile to the extracted spectrum accounting for the instrumental spectral resolution derived from nearby sky lines, using the IDL LINEFIT code (F{\"o}rster Schreiber et al. 2009; Davies et al. 2011).  In the fitting process, the continuum is determined as the mean between the 40th and 60th percentile of the flux level around the \Ha\ and [NII] emission lines {(i.e. in the wavelength range {$\lambda_{H\alpha} \pm 0.02 - 0.05$ $\mu m$, where $\lambda_{H\alpha}$ denotes the wavelength of the redshifted \Ha\ emission line)}}. The fits are performed with Gaussian weighting based on an input noise cube, {and uncertainties in rotation velocity are derived based on a Monte-Carlo (MC) technique, where the input spectrum is perturbed according to the rms noise from the noise cube associated with each data cube.}  Once we have constructed a \rc\ for each galaxy, we determine its observed amplitude and extent by fitting a Freeman exponential disk model of the form:

\begin{equation}
 v_{\rm{disk}}(r) =  \frac{r}{r_d}\sqrt{\pi G\Sigma_{0}r_d[I_0K_0 - I_1K_1]},
\label{freeman.eq} 
\end{equation}

\vspace{01mm}

where $r_d$ is the radial scale-length of the exponential disk, corresponding to $r_d = R_{\rm{e}}/1.68$.  $\Sigma_0 $ is the central mass surface density of the disk, and $I_n$ and $K_n$ denote the modified Bessel functions of the first and second kind (Freeman 1970).  In the fitting, we leave $r_d$ and $\Sigma_{0}$ as free parameters.  {Before fitting, the Freeman disk model is convolved with a 1D Gaussian of FWHM corresponding to the PSF associated with each galaxy.  After the fit has converged, we determine the maximum observed velocity (\vrot) and the radius of the peak (\rturn) from the model.  In the right panels of Figure\ \ref{apertures.fig}, \rcs\ of two examples are shown together with the respective fit.}  

The model represented by Equation\ \ref{freeman.eq} {correctly describes} a disk with infinitely small scale height $h_z$.  As disks at higher redshift are found to be geometrically thicker than their local counterparts based on both their shapes and kinematics (e.g. F{\"o}rster Schreiber et al. 2009; Kassin et al. 2012; van der Wel et al. 2014b; Wisnioski et al. 2015), we {also use} model rotation curves that take into account finite scale heights (Noordermeer 2008).  Assuming a thickness parameter (i.e. ratio of scale height to scale length, $h_z/r_d$) of 0.2, we find that the resulting rotation curves peak intrinsically at larger radii and lower peak rotation velocities (by $6\%$ and $8\%$, respectively) compared to the infinitely thin disk case at fixed $r_d$ and $\Sigma_0$.  However, we find that the inferred \vrot\ and \rturn\ parameters do not change by more than $\sim 1 \%$ in the fitting process when adopting a disk of finite thickness.\\

Typical rotation curves for individual objects within our sample can be constrained out to $\sim R_{\rm{turn}}$ with only a few exceptional high quality RCs among our sample reaching out to several effective radii beyond \rturn.  Those high-quality datasets {have been} presented and analyzed in a separate paper {(Genzel et al. 2017)}.  Therefore, a judgment whether the apparent flattening of a rotation curve results in a flat asymptotic behavior or a turn-over with an outer drop in rotation velocity is not possible for the average galaxy in our sample.  However, we use the above Freeman disk model to quantify \rturn\ as the radius where a turn-over of velocity would be \textit{expected} in the context of the model.  We refer to this turn-over radius as quantified on the observed \rcs\ as `\rturno'. Note that the maximum observed velocity $V_{\rm{max}}$ does not correspond to the intrinsic {peak} rotation velocity ($V_{\rm{rot}}$) since it is not further corrected for inclination, which is not necessary for our 
analysis.\\

To leverage the faint outer emission in regions beyond $R_{\rm{turn}}$ accessible within the FOV of KMOS and SINFONI, we have developed a stacking approach that combines the full spectral information on the rotation curve for each galaxy.  This methodology is based on position-velocity (pv) diagrams {that} are normalized in spectral and spatial directions.  To create those, the extracted spectra are re-sampled onto a common spatial and spectral grid such that each normalized pv diagram has a fixed pixel scale of $0.15 V_{\rm{max}}$ in spectral and $0.16  R_{\rm{turn}}$ in spatial direction.  This corresponds to an oversampling factor of $\sim 1.8$ and $\sim 2.3$ compared to the data cubes, respectively.  We choose these pixel scales as a best compromise between accuracy in the alignment of normalized pv diagrams and resulting $S/N$.  Next, we normalize the surface brightness of each galaxy in the outer regions by scaling each pv diagram by the peak \Ha\ intensity at \rturn.  Through statistical  {tests explained in Section 3.2, this scheme} of flux normalization does not lead to a single object dominating the stack at large radii.\\

\subsection{Final stacking}

\begin {figure}[t]
\centering

 \includegraphics[width=0.48\textwidth]{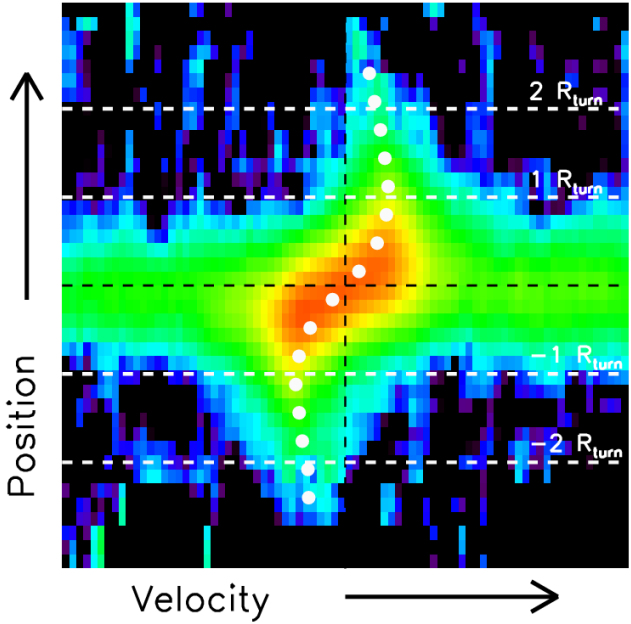}
\caption[Stacked pv diagram]{Final stacked pv diagram shown in logarithmic color scaling.  The zero positions in velocity and radius are marked as black lines, and the radial position of $1R_{\rm{turn}}$ and $2 R_{\rm{turn}}$ are marked as white dashed horizontal lines.  The {white} symbols represent the best-fit velocity determined from Gaussian fits to the extracted spectrum at a given position to outline the shape of the two-sided stacked rotation curve before folding both sides together.}
\label{pv.fig} 
 \vspace{3mm}
\end {figure}

 {We derive the average pv diagram} by computing the mean of all normalized individual pvs.  We note that we obtain consistent results (i.e. a consistent negative outer slope of the stacked rotation curve presented below within the uncertainties) when deriving the stacked pv diagram from the median of all normalized pvs.
Before averaging, pixels that severely suffer from residuals of OH skylines or other sources of noise are discarded. 
 {Figure\ \ref{pv.fig} plots the final stacked pv diagram, with the wavelength (or velocity) axis in horizontal and radial axis in vertical direction. The stacked pv diagram allows tracing the projected rotation curve to about {$2.4 R_{\rm{turn}}$}, as indicated in Figure\ \ref{pv.fig}.}  Beyond $R_{\rm{turn}}$ the resulting rotation velocity is decreasing symmetrically on both sides of the galaxy stack.

To further increase the $S/N$,  {we combine both sides of the stacked pv} by folding together the stacked pv diagram around the central spatial and spectral axes.  On this folded pv diagram, the final stacked \rc\ is then derived by performing Gaussian fits with the LINEFIT code determining the centroid on the averaged \Ha\ emission line at different radial positions.  At each radius, the spectra of two {adjacent} spatial bins are summed up prior to extracting the velocities.  Prior to fitting, the spectra are median-filtered with a kernel of five normalized spectral pixels {(corresponding to a spectral window of {$0.75 \,V/V_{\rm{max}}$})} to smooth out `wiggles' in the resulting \Ha\ line profile originating from stacking emission lines with slightly different centroid positions (due to small centroiding errors  {and} intrinsic variance of rotation curve shapes among the stacking sample).  We verified that both re-sampling the pv into radial bins of two pixels prior to extraction as well as additional spectral 
median-filtering do not change the shape and inferred slopes of the outer stacks on a level significant compared to the uncertainties.  However, omitting either of these two steps leads to slightly larger uncertainties (up to $\sim 20 \%$) in the inferred outer slopes.\\

 {We derive the velocity errors of the stacked RC using a bootstrapping technique that encompasses sample variance and the statistical pixel-to-pixel RMS in the spectra. We iterate 300 times,} each time drawing a random sample from the original dataset allowing for replacement.  In every bootstrap iteration, we perturb each pv by its pixel-to-pixel noise before stacking.  The 68\% scatter in the distribution of all bootstrapped iterations is taken as the error in the stacked RC.  Overall, we find that the uncertainties are roughly to equal extent driven by statistical RMS noise in the stacked spectra and by the sample variance.\\

Due to the pv normalization in  {the} radial direction, each normalized pv diagram extends out to a certain radius depending on the value of \rturn\ relative to the FOV for a given galaxy.  This translates into  {fewer} galaxies that contribute to the final stack with increasing galactocentric radius.  We therefore additionally employ a jackknifing technique where we repeat the stacking, each time removing one galaxy from the sample, and verify that the stack in the outer part is not dominated by a single object.  We find that this holds for galactocentric radii $ \lesssim 2.4 R_{\rm{turn}}$, where the number of contributing galaxies is larger than 10.\\
Furthermore, the masking of contaminating noisy pixels before averaging causes the effective number of galaxies that are averaged into our stack at a given spectral and spatial position to be lower than the total number of galaxies in our sample.  We account for this effect by only counting galaxies where more than 50\% of the spectrum is not masked out {within { $\pm \sim4 \, V/V_{\rm{max}}$}, which corresponds to the normalized wavelength region we use to constrain the Gaussian fits to derive the velocity measurements on the stack.}  In the remainder of this work, we refer to this number as the `effective number of galaxies' contributing as a function of radius.\\

\subsection{An alternative calibration of \rturn}
\label{rturn.sec}

\begin {figure*}[t]
\centering
 \includegraphics[width=0.72\textwidth]{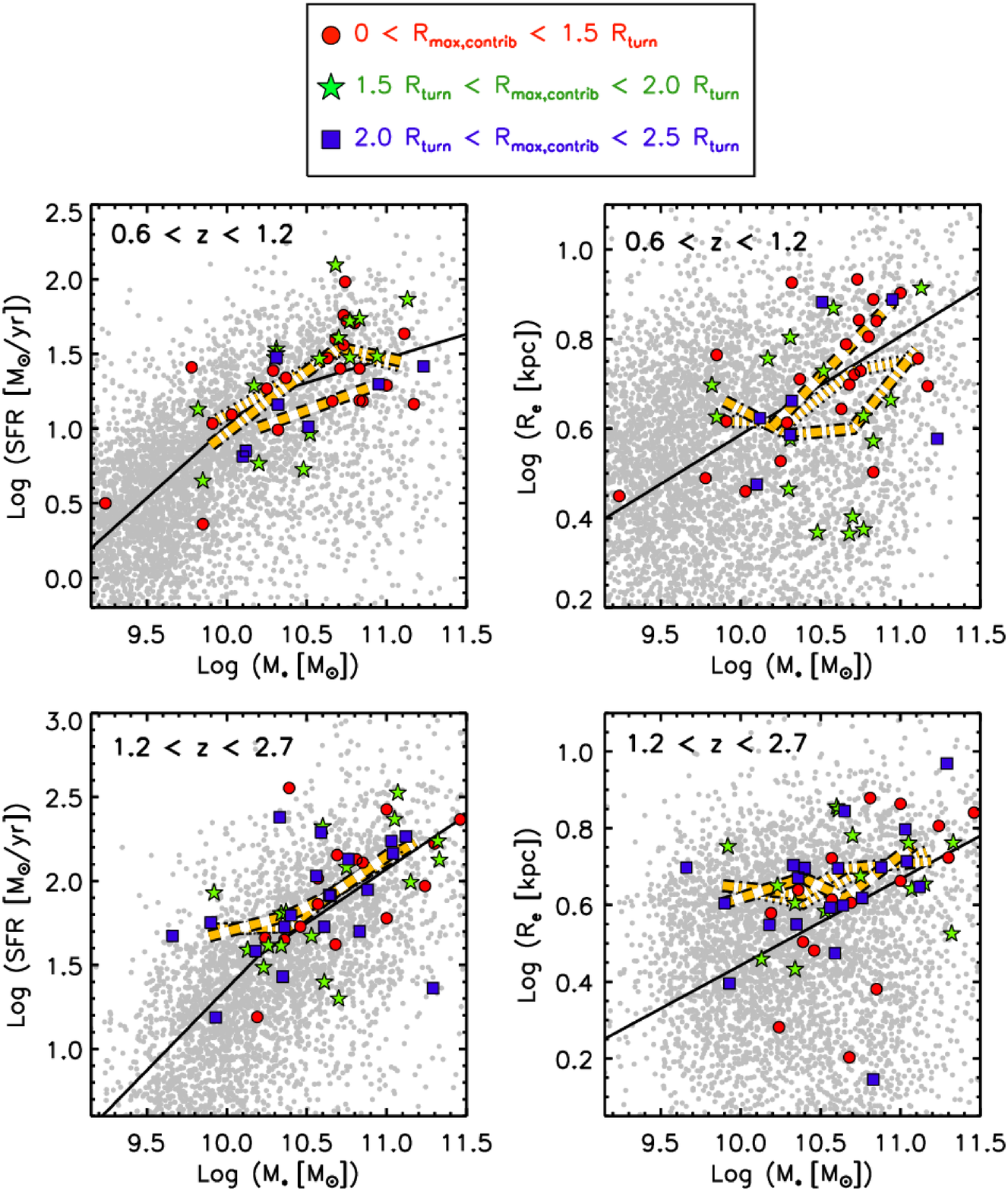}
\caption[Stacked sample in the $R_{\rm{e}} - M_*$, $SFR - M_*$ planes for different radii]{Properties of the stacked galaxy sample in the $SFR - M_*$  plane (left) and $R_{\rm{e}} - M_*$ plane (right).  {{ In each panel, the colored symbols represent galaxies that contribute to radii $0 < R_{\rm{max,contrib}} < 1.5$ \rturno\ (red circles), $1.5$ \rturno\ $ < R_{\rm{max,contrib}} < 2$ \rturno\ (green stars) and $2$ \rturno\ $ < R_{\rm{max,contrib}} < 2.5$ \rturno\ (blue squares).  Additionally, we show running medians as thick orange lines representing the different groups ($ R_{\rm{max,contrib}} > 0$ \rturno\ as dotted lines, $ R_{\rm{max,contrib}} > 1.5$ \rturno\ as dotted-dashed lines, and $ R_{\rm{max,contrib}} > 2$ \rturno\ as dashed lines). }}
The underlying 3D-HST population is represented by grey symbols. The black solid lines in the left panels represent the broken-power law parametrization of the main sequence by Whitaker et al. (2014) for the corresponding redshifts.  The solid lines in the right panels represent the size-mass relation of late-type galaxies derived from van der Wel et al. (2014a).}
\label{sample_properties3.fig}
 \vspace{4mm}
\end {figure*}

The fit to determine \rturno\ and \vrot\ relies on the assumption that the shape of the baryonic mass profile is exponential, which we have validated based on the $H$-band \sersic\ index distribution for our sample (see Appendix A).  However, central stellar bulges, most pronounced in the stellar mass distributions, are found to be significant in the most massive SFGs at $0.5 < z < 2.5$, {with bulge mass fractions reaching up to {$\sim 40-50 \%$} at  {$\log{(M_*/M_{\sun})} > 11$}} (Lang et al. 2014).  {The shape of the inner rotation curve of galaxies (and in particular the position of the turn-over relative to the intrinsic \Re) depends on details of the inner mass distribution in galaxies (e.g. Kent 1986; Corradi \& Capaccioli 1990; Noordermeer et al. 2007; Noordermeer 2008), as well as on the convolution of the projected velocity map with the beam size.}  {We therefore} employ an alternative method to determine \rturn.  This technique converts a measurement of the intrinsic effective radius $R_{\rm{e}}$ to an `observed' turn-over radius (hereafter referred to as `\rturnm'),  {taking into account the \sersic\ index, galaxy inclination, and the amount of beam smearing that affects the observed \rc}. Values for $R_{\rm{e}}$, the \sersic\ index and the inclination are derived from the available $H$-band HST imaging assuming that the observed $H$-band represents the extent and shape of the baryonic distribution (see Section\ \ref{ancillary.sec}).\\ 

Our sample shows good correspondence between the \rturno\ and \rturnm\ {(see top right panel of Figure B1, and discussion in Appendix B)}, demonstrating the robustness of our \rturno\ determination using the Freeman disk model.  The details of this conversion and the resulting \rturnm\ are presented in Appendix B, together with a discussion about the impact of using the \rturnm\ on the final results presented in the remainder of this paper.

\vspace{0.5cm}

\subsection{Sample properties at different galactocentric radii}

Here, we demonstrate that the  {decreasing number of galaxies in the stack} does not result in large variations in the average galaxy parameters examined in Section 2.3 with increasing galactocentric radius.  {We split our entire galaxy sample into different groups depending on the maximum radius to which they are contributing in the stack ({$R_{\rm{max,contrib}}$} ) and examine their location in the $SFR - M_*$ and $R_{\rm{e}} - M_*$ planes as illustrated in Figure\ \ref{sample_properties3.fig}.} 

Considering the $SFR - M_*$ plane, galaxies that contribute to the middle and outer regions of the stack still cover homogeneously the main sequence over the entire redshift range.  At $z < 1.2$, the median SFRs of the blue points are offset by $\sim 0.3 - 0.4$ dex with respect to the full sample, which is, however, smaller than or comparable to the scatter among the blue points.  Therefore, we conclude that galaxies probing the middle and outer parts of the stack are still representative of the MS population over our entire redshift range.\\ 
Turning to the size distributions, we also find good overlap among galaxies contributing to the middle and outer parts of the stack compared to the entire stacking sample.  We also find only little variations in the median \sersic\ indices for galaxies contributing to different radial ranges ($\Delta(n) \sim 0.3$), such that those are in good agreement with exponential distributions at all radii in the stack.

\section {Results}
\label{structure.sec}

\subsection{Shape of the stacked rotation curve}
\label{shape.sec}

\begin {figure*}[htbp]
\centering

 \includegraphics[width=0.65\textwidth]{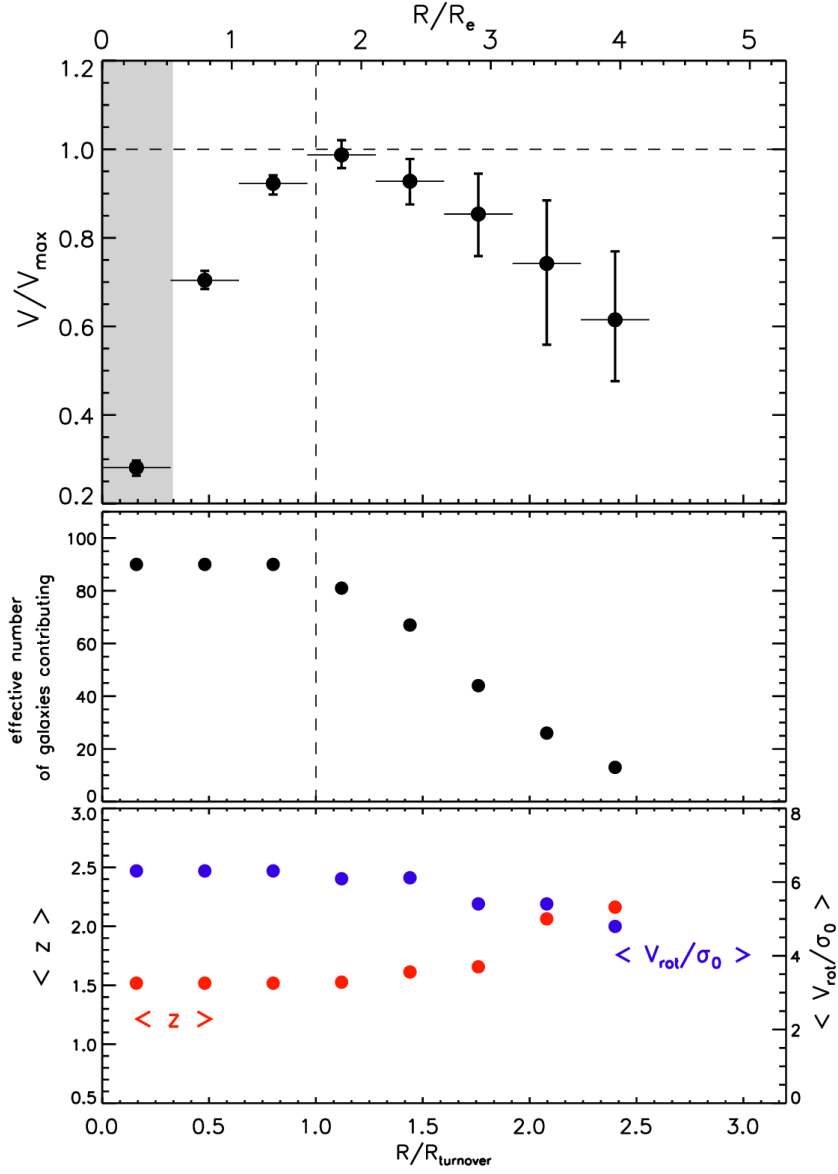}
\caption[Stacked rotation curve]{Top: Stacked rotation curve (black dots) plotted in units of normalized velocity ($V/V_{\rm{max}}$), normalized radius ($R/R_{\rm{turn}}$), and intrinsic effective radius ($R/R_{\rm{e}}$).  The error bars are derived from bootstrapping and include both sample variance as well as RMS noise in the spectra. The shaded area marks the half-light beam size of the average PSF observed for our sample. 
Middle: Effective number of galaxies contributing to the stack, accounting for masking out noisy pixels in the pv diagrams. The decrease in the number of contributing galaxies with increasing radius is driven by FOV limitations. Bottom: Median redshift and $V_{\rm{rot}}/\sigma_0$ of contributing galaxies for a given radial bin.}
\label{rc.fig}
 \vspace{3mm}
\end {figure*}

Figure\ \ref{rc.fig} displays the best-fit velocity measurements on the stacked and folded pv diagram, shown in the {{normalized coordinate frame [$V/V_{\rm{max}}$,$R/R_{\rm{turn}}$].  Here and in the remainder of this paper, we quote the observed and normalized rotation velocity of our stack in units of '$V/V_{\rm{max}}$'.}}  We furthermore convert $R/R_{\rm{turn}}$ into units of intrinsic $R_{\rm{e}}$ on the basis of an exponential rotating disk as well as  accounting for the average spatial beam smearing, and obtain the relation $R_{\rm{turn}}/R_{e} \sim 1.65$.  The error bars of the stacked \rc\ represent the 68\% scatter in the distribution of bootstraps around the median and include both sample variance  {and} spectral RMS noise.  The middle panel indicates the effective number of galaxies that contribute to each velocity bin.  {In the bottom panel we show} the median redshift and $V_{\rm{rot}}/\sigma_0$ values of all galaxies contributing at a given radius, since we find that the median of these parameters {notably} change with radius.  {{ Specifically, we find that the median redshift of galaxies contributing to each radial bin smoothly increases with radius, up to $z \sim 2.2 $ beyond  $R > 2 R_{\rm{turn}}$.  We attribute the evolution of decreasing $R_e$ (and thus decreasing \rturn) with redshift as potential cause for this effect.  Analogously, the median $V_{\rm{rot}}/\sigma_0$ of contributing galaxies decreases with radius by up to $\sim 30\%$ for the outermost radial bin, consistent with the above trend via the positive correlation of $\sigma_0$ with redshift (e.g. Kassin et al. 2012; Wisnioski et al. 2015).}}\\

Beyond the inner rising part, the stacked \rc\ exhibits a turn-over and an outer decrease of rotation velocity, also seen as a symmetric fall-off on either side from the center based on our stacked pv diagram (see Figure\ \ref{pv.fig}).  The drop in velocity reaches down to { {$ \sim 62 \%$ of the maximum normalized velocity $V_{\rm{max}}$ at $\sim 2.4 R_{\rm{turn}}$}}. To evaluate the significance of the drop in velocity beyond \rturn\, we derive a linear slope by fitting the five outer velocity bins, inverse variance-weighted by their respective errors, and find a negative slope of $\Delta V / \Delta R = -0.26^{+0.10}_{-0.09}$.  The quantities $\Delta V$ and $\Delta R$ are computed in units of normalized coordinates $V/V_{\rm{max}}$ and $R/R_{\rm{turn}}$, respectively.  In the remainder of this work, we quote all outer slopes $\Delta V / \Delta R$ in this normalized unit, which we refer to as `$[V/V_{\rm{max}}$,$R/R_{\rm{turn}}]$'. 
The uncertainty represents the scatter of slopes determined from the bootstrapping technique discussed in Section\ \ref{method.sec}, and accounts for possible correlations among radial bins.  Also based on our bootstrapping technique, {we find that our outer stacked rotation is falling with respect to a flat behavior at the $99.4 \%$ level.}

We find that our outer stacked \rc\ is in good agreement with both a pure baryonic thin exponential disk as well as an outer Keplerian fall-off (both predicting $\Delta V / \Delta R \sim -0.24\, [V/V_{\rm{max}}$,$R/R_{\rm{turn}}]$, accounting for the average spatial resolution of our data), to first order with only a weak or no imprint of an additional dark matter component within the radial range probed.  However, there are additional mechanisms such as pressure support in the outer disk  {that} need to be considered, and  {that} are discussed together with a more detailed comparison of our stack to modeled rotation curves including baryons and dark matter {below.} \\

 {As discussed in the previous section, we repeat our stacking process using {\rturnm\ }for each galaxy in our sample}.  The resulting stacked rotation curve is displayed in the bottom of Figure B1 as red symbols and compared to the stack as presented in Figure\ \ref{rc.fig} as black symbols.  Despite the larger uncertainties of the \rturnm\ - normalized stack, we measure {an outer slope consistent with that of our nominal stack}  in Figure\ \ref{rc.fig} within the uncertainties.  In the remainder of this paper, we will adopt the stacked rotation curve presented in Figure\ \ref{rc.fig} as the fiducial rotation curve, which we will compare to models.\\

{In this section we have shown that our stacking approach is able to constrain an extended rotation curve representative of our stacking sample, reaching out to $R = 2.4 R_{\rm{turn}}$.}  We observe a rapid, symmetrical fall-off in agreement with a thin exponential disk or {Keplerian} fall-off within the radial range probed.  In the following sections, we demonstrate that this fall-off deviates {at $> 3 \sigma$ significance} from the average rotation curves of local spirals.  {We then compare} our stack with combined baryonic and dark matter components to interpret our findings in view of {mass fractions of baryons relative to dark matter} of massive high-z SFGs.
\subsection{Comparison to rotation curves of local spirals}
\label{catinella.sec}

 {For comparison with local spirals, we} use the compilation of template \rcs\ from Catinella et al. (2006), who determined averaged \rcs\ of a large sample of $\sim 2200$ low-z disks, based on long-slit \Ha\ observations.  Their template averages are parameterized in bins of absolute $I$-band magnitude and are sampled at galactocentric radii comparable to our stack.\\

\begin {figure}[t]
\centering
  \includegraphics[width=0.44\textwidth]{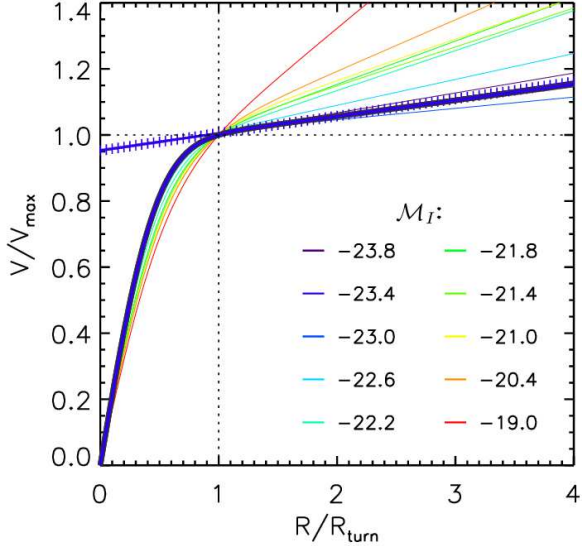}
\caption[Template RCs at low redshift]{Low redshift template rotation curves from Catinella et al. (2006) in different bins of absolute $I$-band magnitude ( Values of ${\rm \mathcal{M}}_I$ according to the center of each bin are indicated in the lower right).  The templates have been beam smeared and normalized in radius and velocity, plotted in the same normalized coordinate frame as Figure\ \ref{rc.fig}.  {The template corresponding to a baryonic mass equal to that of our stacked sample is plotted as thick blue continuous line, with the fitted outer slope overplotted as straight line with `+' symbols.}}
\label{template_rcs.fig}
 \vspace{3mm}
\end {figure}

First, we create low-z template RCs as a function of $R/r_d$ for each bin in absolute $I$-band magnitude (${\rm \mathcal{M}}_I$) using the Polyex parametrization as found in Table 1 in Catinella et al. (2006).  We convolve each RC with a Gaussian beam of FWHM representative for our sample to bring the low-z template RCs to the same spatial resolution as our stacked RC, since the template RCs in Catinella et al. (2006) are well resolved.  The smeared \rcs\ are converted into the same frame of normalized coordinates [$V/V_{\rm{max}}$,$R/R_{\rm{turn}}$].  {Since no peak in the template \rcs\ is observed, we estimate the position of the turn-over radius as expected from a pure baryonic disk (peaking intrinsically at $R/r_d = 2.15$), where the same beam smearing has been applied.}  Finally, we determine the linear outer slope on the convolved and normalized template \rcs\ at galactocentric radii out to $\sim 2.4 $\rturn\ as in the case of our stack.  The convolved and 
normalized template rotation curves are presented in Figure\ \ref{template_rcs.fig}, plotted in the same {normalized} coordinate frame as our stack presented in Figure\ \ref{rc.fig}.  

To choose an appropriate comparison template corresponding to a massive disk with a baryonic mass equal to the median of our stacked sample ($<\log(M_{\rm{baryonic}})> = 10.87 $), we compute the expected $I$-band magnitude for such a disk at low redshift.  Assuming a $(M_*/L)_I \sim 1$, based on population synthesis models from Bell \& de Jong (2001) and average galaxy colors of disks ($B-V \sim 0.5$, e.g. Buta et al. 1994), we find that the average template \rc\ of ${\rm \mathcal{M}}_I = -23.4$ {(highlighted in Figure\ \ref{template_rcs.fig})} yields a good low-z comparison case to our data at the same baryonic mass.  The measured outer slope for the corresponding normalized template RC gives a positive value of $\Delta V / \Delta R \approx 0.06\, [V/V_{\rm{max}}$,$R/R_{\rm{turn}}]$.  This value deviates from our \textit{falling} stacked \rc\ with a significance level of $\sim 99.9$ \% (corresponding to $\sim 3.2 \sigma$, with $\sigma$ denoting the 68\% scatter in the distribution of slopes obtained from our bootstrapping, {neglecting the small uncertainty in slope of the highlighted template \rc )}.\\
Thus, the outer slope of our stack differs from the positive slope of local spirals at similar mass, spatial resolution and radial range on a statistically significant level.  We note that our stacked rotation curve also falls-off more steeply than the rare cases of observed declining rotation curves in some local spirals (e.g. Casertano \& van Gorkom 1991; Dicaire et al. 2008; Noordermeer et al. 2008).\\
There might be additional kinematical effects such as warping, perturbations by minor merging, or radial motions  {that} are capable of producing the outer fall-off observed in our stack.  Warps are frequently observed in the outer HI layers (e.g. Bosma 1978; Briggs 1990) as well as in the optical stellar disks (e.g. S\'{a}nchez-Saavedra et al. 1990) of local disk galaxies, mostly found as point-symmetric S-shaped features, such that one side of the plane of the disk rises and the other side declines.  However, the elevated level of velocity dispersions of high-z SFGs compared to their local counterparts 
suggest an increased stability against warping and buckling instabilities.\\  

\subsection {Comparison with baryonic plus dark matter models}
\label{models.sec}

\subsubsection{The rotating disk + halo model}

In order to compare our stacked RC with models, we simulate a rotating baryonic disk embedded in a DM halo.  For this purpose, we assume an {exponential baryonic disk (thickness = 0.2 ), appropriate for high-z disks.}  The expected rotation curve for such a disk, $v_{\rm{disk}}(r)$,  is computed following Noordermeer (2008).  {The redshift, effective radius and total baryonic disk mass is set} by the median properties of our sample (see Table 1).

\vspace{2mm}

 {We assume {a} Navarro-Frenk-White (NFW) profile} with the density distribution (Navarro et al. 1997): 
\begin{equation}
 \rho(r) = \frac{\rho_0}{(r/r_s)(1+r/r_s)^2}, 
\end{equation}

where $\rho_0$ is the central density of the halo, and $r_s$ is the scale radius.  It is related to the halo's virial radius, $r_{200}$ via $r_s = r_{200}/c$  where $c$ is the concentration parameter. 

The rotation curve of such a dark matter halo is then given by

\begin{equation}
\frac{v^2_{DM}(r)}{V^{2}_{200}}= \frac{r_{200}}{r}\frac{\ln(1+r/r_s)-(r/r_s)/(1+r/r_s)}{\ln(1+c)-c/(1+c)}.
\end{equation}

The halo's virial velocity $V_{200}$ and virial mass $M_{200}$ are connected through $r_{200}$ in the following way (Mo et al. 1998):
\begin{equation}
r_{200}(z) = \frac{V_{200}(z)}{10H(z)} ,\, M_{200} = \frac{V^3_{200}(z)}{10GH(z)}.
\end{equation}
For a flat $\Lambda$CDM universe, the Hubble parameter $H(z)$ is connected to $H_0$ via 

\begin{equation}
 H(z) = H_0 \sqrt{\Omega_{\Lambda,0} + \Omega_m \cdotp (1+z)^3 }.
\end{equation}

Given these relations, the rotation curve of a NFW halo is completely defined by setting a virial mass, concentration parameter and redshift.  We compute $M_{200}$ by setting $m_{\rm{d}}$, the mass fraction of the baryonic disk over the halo within $r_{200}$, given by $m_{\rm{d}} = \frac{M_{\rm{baryonic}}} {M_{200}}$.  The final model rotation curve is then given by the circular velocity $v_{\rm{circ}}(r)$, computed by combining the baryonic plus dark matter component :

\begin{equation}
v_{\rm{circ}}^2(r)= v_{\rm{disk}}^2(r) + v_{\rm{DM}}^2(r).
\label{vel_tot.eq}
\end{equation}

To make appropriate comparisons of our modeled rotation curves with the data, we convolve our model with a Gaussian PSF of FWHM representing the average of our sample.  {{ To convert the model into the observed units $V/V_{\rm{max}}$ and $R/R_{\rm{turn}}$}, we scale the model} in velocity and radius such that the turn-over in the model is [1,1] in the coordinate frame of [$R/R_{\rm{turn}}$,$V/V_{\rm{max}}$].

Numerical simulations show that the concentration of halos, $c$, is anti-correlated with halo mass and decreases with redshift (e.g. Bullock et al. 2001; Zhao et al 2009; Prada et al. 2012). 
For our comparison we compute $c$ on the basis of the parametrization given by Bullock et al. (2001):

\begin{equation}
c = 9\mu^{-0.13},
\label{Bullock.eq}
\end{equation}

where $\mu = M_{200}/(1.5 \cdotp 10^{13} h^{-1} M_{\sun})$.  This equation is valid for $z=0$, and $c$ has a redshift dependence of $c \propto (1 + z)^{-1}$.  Due to the mass dependence, we compute $c$ for each modeled rotation curve depending on halo mass (resulting from setting the $m_{\rm{d}}$ parameter).  {For the range of halo masses considered here { ($\log{M_{200}/M_{\sun}} \sim 11.6 - 12.4 $)}}, we find that $c$ varies roughly from 4 to 7.  However, we verify that further variations in the concentration parameter only marginally affect our models presented below by exploring a large range of $c$ from 2 to 12, discussed in Section\ \ref{dependencies.sec}. \\

In our modeling, we fold in uncertainties in the mass scaling of the baryonic disk by perturbing $\log(M_{\rm{baryonic}})$ by $\sim 0.3$ dex, reflecting the systematic errors in population synthesis modeling (with the most dominant error source arising from assumptions on the IMF, see e.g. Kauffmann et al. 2003) as well as uncertainties in the inferred gas masses.  All other model parameters are held fixed in this process, {such that the model error budget solely reflects the uncertainty in the total baryonic mass.}

\subsubsection{Correction for pressure support}

Since high-z disks are observed to exhibit turbulent gas motions with {large velocity dispersions} (e.g F{\"o}rster Schreiber et al. 2009; Kassin et al. 2012; Wisnioski et al. 2015), we furthermore take into account the effect of pressure support to the disk.  Assuming a hydrostatic pressurized gas disk with an exponential scale length $r_d$, and an intrinsic velocity dispersion $\sigma_0$, the observed rotation velocity $v_{\rm{rot}}$ is given by (Burkert et al. 2010):

\begin{equation}
 v_{\rm{rot}}^2(r) =  v_{\rm{circ}}^2(r) - 2\sigma_0^2 \Big(\frac{r}{r_d}\Big),
\label{pressure.eq}
\end{equation}

where the velocity $v_{\rm{circ}}(r)$ is the circular velocity, i.e. the rotation velocity in the case of no pressure support.  Note that Equation\ \ref{pressure.eq} assumes an exponential mass distribution.  For systems with non-negligible $\sigma_0$ compared to the inclination-corrected rotation velocity $V_{\rm{rot}}$, $v_{\rm{rot}}(r)$ is reduced significantly beyond \rturn, which can lead to a decline of rotation velocity steeper than Keplerian {( $v_{\rm{rot}}(r) \sim \sqrt{1/r}$ ) in the outer disk.}

Equation\ \ref{pressure.eq} is applied to our models via setting $V_{\rm{rot}}/\sigma_0$, where $V_{\rm{rot}}$ is the intrinsic rotation velocity at the peak of the rotation curve that is easily determined from the model.  
Due to the decrease of $V_{\rm{rot}}/\sigma_0$ with radius, ranging from $\sim 6.3$ to $\sim 4.8$ for the outermost radial bin, we adopt for a given radius a model with $V_{\rm{rot}}/\sigma_0$  according to the  profile shown in the bottom panel of Figure\ \ref{rc.fig}, such that the effective value of $V_{\rm{rot}}/\sigma_0$ is decreasing with radius.

\subsubsection{Predicted rotation curves}

\begin {figure*}[htbp]
\centering
 \includegraphics[width=0.64\textwidth]{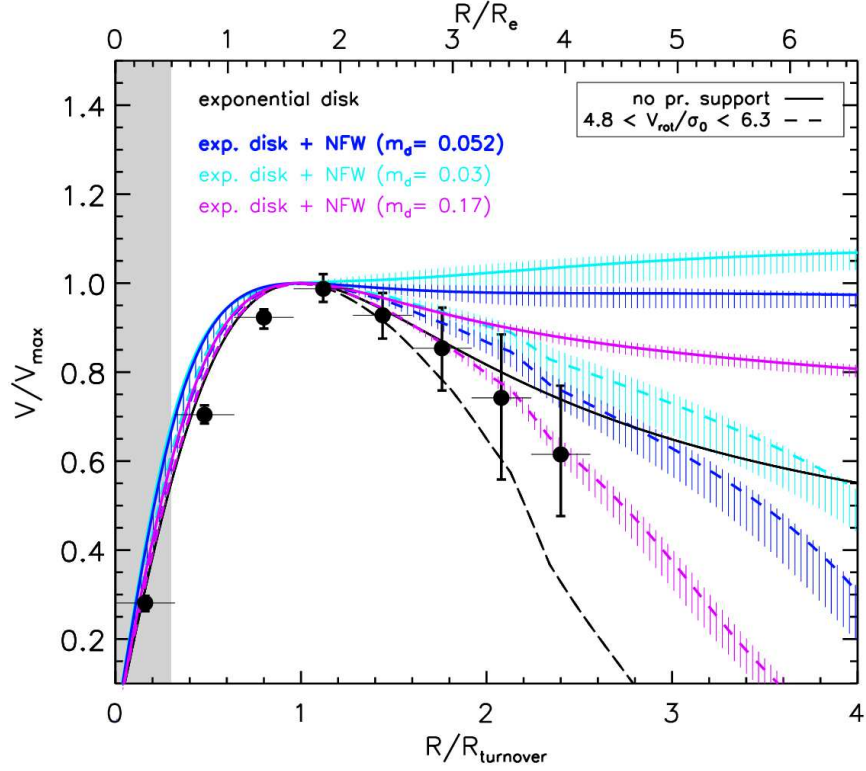}
\caption{Normalized stacked rotation curve (black circles) shown together with a baryonic-only exponential disk model (black lines), and models with added dark matter halo components for different baryonic mass fraction in the disk, $m_{\rm{d}}$ (colored lines). Each model is convolved with a PSF of the representative average FWHM of the sample before normalization. Also shown are versions of each models with implemented pressure support (dashed lines) with $V_{\rm{rot}}/\sigma_0$ adopted to decrease with radius according to Figure\ \ref{rc.fig}, ranging from $6.3$ to $4.8$.}
\label{model_rc.fig}
 \vspace{4mm}
\end {figure*}

We now explore our models with different baryonic disk mass fractions, $m_{\rm{d}}$, first neglecting pressure effects.  In Figure\ \ref{model_rc.fig} we plot the stacked RC together with a normalized pure exponential disk model and additional baryonic plus dark matter models with a range of baryonic disk fractions $m_{\rm{d}}$ as solid colored lines.  The pure exponential disk without any additional dark matter already yields a good fit to the data as pointed out in Section 4.1. 

In order to consider a fiducial model including dark matter, appropriate for high-z disk galaxies, we adopt $m_{\rm{d}} = 0.052$ from Burkert et al. (2016).  Their estimate of $m_{\rm{d}}$ is based on a large sample of $\sim 360$ massive ($\log(M_*) \sim 9.3 - 11.8$) SFGs galaxies at $0.8 < z < 2.6$, including almost our entire stacking sample used here.  Taking the significant contribution of molecular gas in the disk into account {(e.g. Tacconi et al. 2010, 2013, 2017; Daddi et al. 2010; Genzel et al. 2015)},  $m_{\rm{d}} = 0.052$ is in reasonable agreement with current abundance matching results (Moster et al. 2013; Behroozi et al. 2013). 

The corresponding model predicts a velocity profile that features a turn-over of velocity with an overall flat behavior beyond the turn-over radius.  The strong fall-off of our stack appears to be inconsistent with this flat behavior.\\ 
Next, we consider a model adopting a baryonic mass fraction appropriate for \textit{local} massive disk galaxies.  At $z\sim0$, abundance matching results estimate a stellar mass fraction between $m_{\rm{d}} \approx 0.025$ (Behroozi et al. 2013) and $m_{\rm{d}} \approx 0.035$ (Moster et al. 2013).  Neglecting the amount of gas in massive local spiral galaxies, we thus set  $m_{\rm{d}} = 0.03$ as a \textit{baryonic} mass fraction.  The corresponding model predicts a similar behavior with a slightly more positive outer slope, which is yet more inconsistent with our data. \\
We furthermore test an extreme situation in which all baryons in the dark matter halo are confined within the disk by setting $m_{\rm{d}}=0.17$, representing the cosmic baryon fraction relative to that of the dark matter ($\Omega_b / \Omega_{DM}$).  Although this assumption represents an upper limit on $m_{\rm{d}}$, the corresponding model still predicts a rotation curve that does not fall off as rapidly as our stack indicates.\\

Next, we compare the same models {but turn on} the effects of turbulent gas pressure in the disk, displayed by dashed lines in Figure\ \ref{model_rc.fig}.  The effect of pressure support strongly decreases the modeled rotation velocity at large radii, and leads to a better agreement with our stacked rotation curve.  The pressure-corrected pure baryonic disk falls more steeply than the stack, consistent with an additional, outer dark matter contribution.  Among the range of our models tested, $m_{\rm{d}} = 0.17$ yields the best match to our observations.  Taking into account the uncertainty of the outer linear slope, a baryonic disk fraction of our fiducial high-z model with $m_{\rm{d}} = 0.052$ is, however, still consistent with our data within the $\sim 1 \sigma$ uncertainties.\\

Our comparison demonstrates that our data {support} high baryonic fractions ($m_{\rm{d}} \gtrsim 0.05$) in conjunction with {the effect of} pressure support in the outer disk to explain the steep fall-off {of} the outer stacked rotation curve.  While this is in agreement with current abundance matching results, even higher baryonic disk fractions (up to the \textit{cosmic} baryon fraction) are preferred.  {These findings are in good agreement with our best individual high-quality rotation curves.  From detailed kinematic modelling of these individual galaxies, {Genzel et al. (2017)} previously inferred similarly high baryonic fractions.}\\
The error budget of our models (shown as shaded areas in Figure\ \ref{model_rc.fig}) is overall small and insignificant compared to the predicted {and} observed outer fall-off.  Since {the model errors reflect} the uncertainties in the adopted baryonic mass, our findings and conclusions made in the context of our models are unaffected by variations in the assumed light-to-mass conversions.\\

{
{Replacing the adopted decrease of $V_{\rm{rot}}/\sigma_0$ with radius by a constant value in our models (as assumed for an individual galaxy) does not alter the major conclusions stated in this section.  Specifically, when setting $V_{\rm{rot}}/\sigma_0 = 6.3$, the predicted model slopes slightly increase, such that the $m_{\rm{d}} = 0.052$ - model including pressure support is consistent with our data at $\sim 1.4 \sigma$ significance, with $m_{\rm{d}}=0.17$ still yielding the formally best fitting model to our data.}}\\

\begin {figure}[tb]
\centering
 \includegraphics[width=0.47\textwidth]{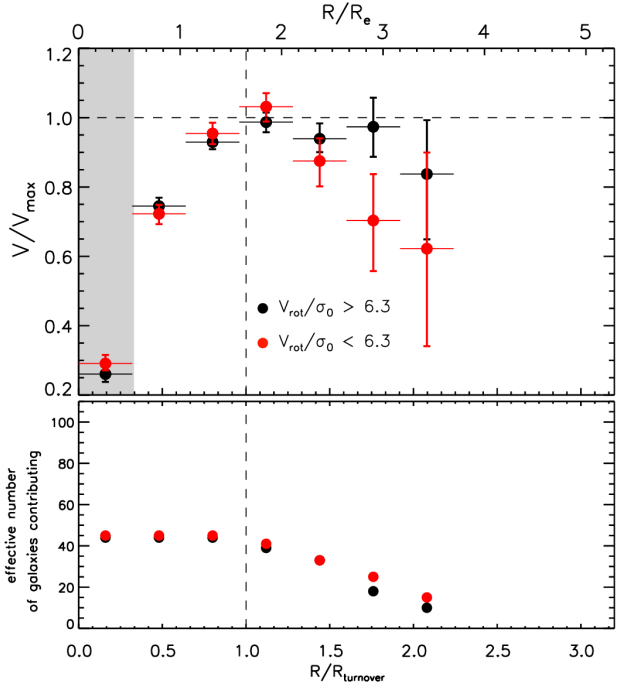}
\caption{Comparison of stacked rotation curves from two subsamples with $V_{\rm{rot}}/ \sigma_{0} < 6.3$ (red symbols) and $V_{\rm{rot}}/ \sigma_{0} > 6.3$ (black symbols).}
 \label{v_sigma_bin.fig}
 \vspace{3mm}
\end {figure}

In order to further substantiate our conclusion that the drop in the outer stack is to a {sizeable} extent driven by pressure effects, we split our sample into two bins {according to their {$V_{\rm{rot}}/\sigma_0$} ratios} and repeat the stacking process for each bin.  The sample is divided around the median $V_{\rm{rot}}/\sigma_0$ of 6.3, such that each bin contains an equal number of galaxies.  The resulting stacked rotation curves are displayed in Figure\ \ref{v_sigma_bin.fig} as black and red symbols, respectively.  The two curves exhibit a different behavior beyond the turn-over radius despite the larger uncertainties due to decreased number of galaxies in each bin.  The 
stack of galaxies with $V_{\rm{rot}}/\sigma_0 > 6.3$ exhibits a flatter behavior with an outer slope of $\Delta V / \Delta R = -0.10^{+0.10}_{-0.11}\, [V/V_{\rm{max}},R/R_{\rm{turn}}]$, compared to the lower $V_{\rm{rot}}/\sigma_0$ stack with a slope of $\Delta V / \Delta R = -0.43^{+0.18}_{-0.17}\, [V/V_{\rm{max}},R/R_{\rm{turn}}]$.  {This confirms the expected trend from Equation\ \ref{pressure.eq}, and supports} the conclusion that {a sizeable part} of the drop seen in the stacked rotation curve is driven by pressure effects in the outer disk.

{
{Examining how the two rotation curves in Figure\ \ref{v_sigma_bin.fig} compare to our baryonic + dark matter models, we find that the stacks with both low and high $V_{\rm{rot}}/\sigma_0$ - galaxies are formally best fit by a model with $m_{\rm{d}}=0.17$, where  $V_{\rm{rot}}/\sigma_0$ is adopted accordingly (i.e. $\sim 4.2$ and $\sim 9.9$ for the low, and high $V_{\rm{rot}}/\sigma_0$ - bin, respectively). As in the case of our comparison above, a model with $m_{\rm{d}}=0.052$ yields an outer slope that is consistent with the data within the $\sim 1 \sigma$ uncertainties.}}

Among various galaxy parameters, we find that, most significantly, the median redshifts of the two $V_{\rm{rot}}/\sigma_0$ bins are different ($\sim 2.2$ and $\sim 0.9$ for the low and high  $V_{\rm{rot}}/\sigma_0$ bin, respectively), consistent with the observed increase of velocity dispersion with redshift (e.g. Kassin et al. 2012; Wisnioski et al. 2015). 

\begin {figure}[hbt]
\centering
 \includegraphics[width=0.44\textwidth]{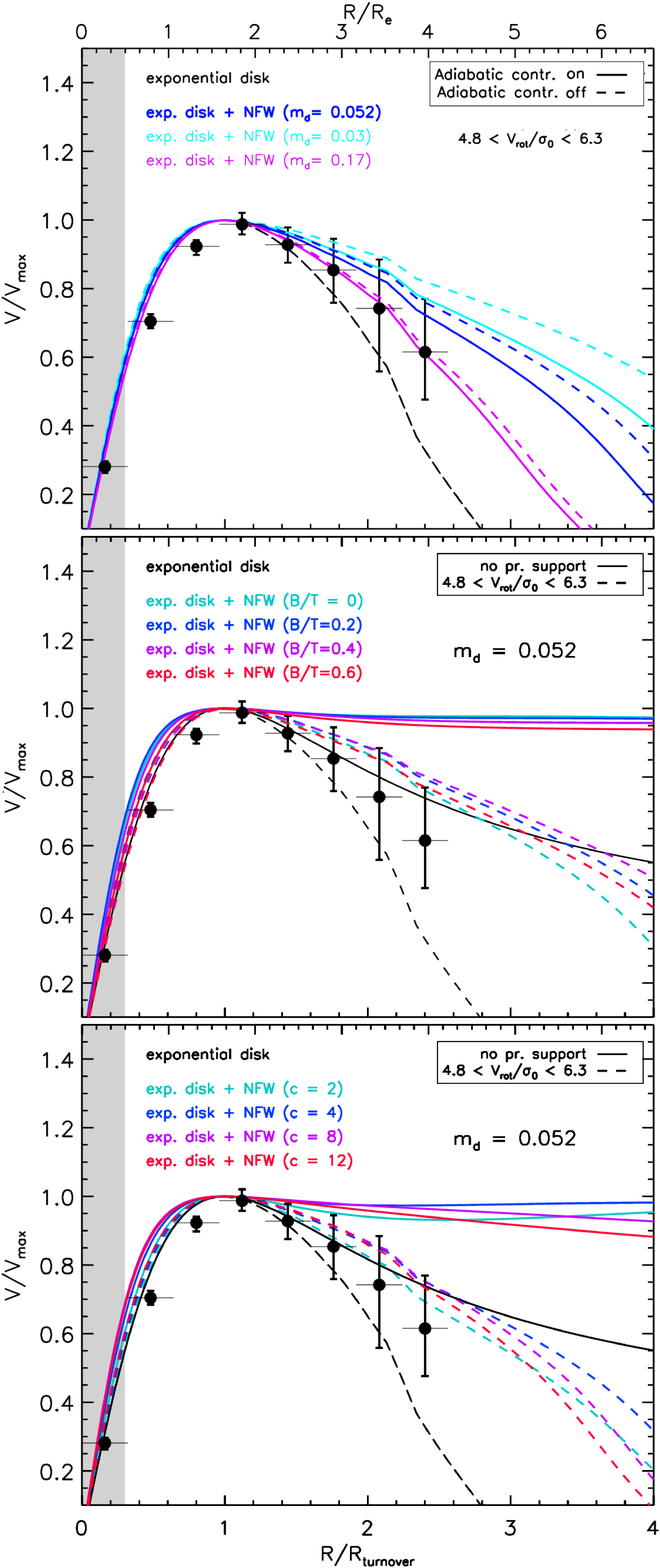}
\caption{Comparison of the stacked rotation curve with the models discussed in Section 4.4 where different parameters have been modified. Top: Models for a range of $m_{\rm{d}}$ including pressure correction ($4.8 < V_{\rm{rot}}/\sigma_0 < 6.3$), shown with and without the effect of adiabatic contraction (solid and dashed lines, respectively). Middle: Models including central bulge components with a range of B/T ratios, adopting $m_{\rm{d}} = 0.052$.  Bottom: Models for a range of concentration parameters, adopting $m_{\rm{d}} = 0.052$. }
 \label{dependencies.fig}
 \vspace{3mm}
\end {figure}

\subsection{Exploring model dependencies}
\label{dependencies.sec}

In this Section, we demonstrate that variations in our model assumptions concerning the effect of adiabatic contraction of the halo, the presence of central stellar bulge components, and variations in the halo concentration parameter (summarized in Figure\ \ref{dependencies.fig}) do not affect the conclusions made in this study.

\subsubsection{Adiabatic contraction}

To test and include adiabatic contraction of the dark matter halo in our models, we adopt the Blumenthal et al. (1986) prescription, which can be arranged to yield the implicit equations (Burkert et al. 2010):

\begin{equation}
v_{\rm{circ}}^2(r) = v_{\rm{disk}}^2(r) + v_{\rm{DM}}^2(r^{'}), 
\label{adi1.eq}
\end{equation}

where 

\begin{equation}
 r^{'} = r\Big(1 + \frac{r \cdotp v^2_{\rm{disk}}(r)}{r^{'} \cdotp v^2_{\rm{DM}}(r^{'})}\Big).
\label{adi2.eq}
\end{equation}

We solve these equations numerically and replace Equation\ \ref{vel_tot.eq} with Equations \ref{adi1.eq}-\ref{adi2.eq}.\\

The top panel of Figure\ \ref{dependencies.fig} plots the resulting models equivalent to the ones discussed in the previous sections, adopting a range of $m_{\rm{d}}$ ratios.  The impact of adiabatic contraction on the normalized modeled rotation curves is small for a given value of $m_{\rm{d}}$ compared to the uncertainties of our data, leading to slightly decreased normalized outer velocity.  We thus conclude that the imprint of adiabatic contraction cannot be tested by our stacked rotation curve since the shape of the rotation curve normalized in velocity changes only on marginal levels.  

 \subsubsection{Deviations from exponential distributions}

Next, we test how the presence of central stellar bulges is reflected upon the expected outer rotation curves in the context of our models, which assume exponential mass  {distributions}.  
As observed in the local universe, early-type disks with significant bulges show declining rotation curves (e.g. Noordermeer et al. 2007).  Also, van Dokkum et al. (2015) found that the reconstructed RC inferred from galaxy-integrated emission line widths of 10 compact massive {SFGs} at $ 2 < z < 2.5$ is in agreement with a declining rotation curve strongly dominated by the dense massive stellar component.\\

We generate a set of additional model rotation curves with an extra central bulge component computed by:

\begin{equation}
v_{\rm{circ}}^2(r) = v_{\rm{disk}}^2(r) + v_{\rm{bulge}}^2(r) + v_{\rm{DM}}^2(r),
\end{equation}

where we calculate $v_{\rm{bulge}}(r)$ using the Noordermeer (2008) prescription for a spherical body (i.e. with a thickness parameter of 1), assuming a \sersic\ law with $n = 4$.  We set $R_{\rm{e,bulge}}$ to be 1 kpc, as found by the bulge-disk decompositions in Lang et al. (2014) as an average size for stellar bulges at $0.5 < z < 2.5$.  For a given bulge-to-total ($B/T$) ratio, we scale the mass of disk and bulge such that the total baryonic mass of the two components remains unchanged.  To properly incorporate the effect of pressure support in the combined bulge+disk models, we generalize Equation\ \ref{pressure.eq} to obtain an expression where mass profiles obey a \sersic\ law ($\Sigma(r) = \Sigma_0 \exp(-b_n (\frac{r}{R_{\rm{e}}})^{1/n})$), given by:

\begin{equation}
 v_{\rm{rot}}^2(r) =  v_{\rm{circ}}^2(r) - 2\sigma_0^2 b_n \Big(\frac{r}{R_{\rm{e}}}\Big)^{1/n}.
\label{pressure_sers.eq}
\end{equation}

The \sersic\ index needed for Equation\ \ref{pressure_sers.eq} is determined from Appendix B in Lang et al. (2014), providing the relationship between $B/T$, $R_{\rm{e,bulge}}/R_{\rm{e,disk}}$ for a bulge+disk model and $n$ for a corresponding single-\sersic\ profile.  {$R_{\rm{e}}$ in Equation\ \ref{pressure_sers.eq} is set equal to the value used for the pure exponential disk model without a bulge component discussed in Section 4.3.1 (see also Table 1)}.\\

The middle panel of Figure\ \ref{dependencies.fig} displays our fiducial model with $m_{\rm{d}} = 0.052$, and with added bulge components of different mass relative to the disk expressed by the $B/T$ ratio.  We test a range of $B/T$ up to 0.6 as observed for the highest mass {SFGs}. 
The models including a central bulge component within the range of $0.2 <  B/T < 0.6$ show overall only small differences compared to a pure disk model considering the uncertainties of our data.  While the intrinsic rotation curve of the bulge component alone ($v_{\rm{bulge}}(r)$) is strongly centrally peaked around $R_{\rm{e}} = 1$ kpc and falls off rapidly in the outer regions, the imprint of its rotation curve is strongly diminished due to the spatial beam smearing applied. Thus, the expected RC of a bulge-dominated galaxy with added dark matter component is hardly distinguishable from the same model without a bulge contribution.  We observe the same behavior when considering our models with pressure support.  Hence, the presence of central bulge components within our galaxy sample does not alter the conclusions made in this work.

 \subsubsection {Variations in the concentration parameter}

Next, we explore how variations in the concentration parameter are reflected by the modeled rotation curves, especially since dark matter halos at lower redshift are expected to be more concentrated (with $c \sim 12$ at $z=0$ and $M_{200} \sim 2 \cdotp 10^{12} M_{\sun}$, e.g. Bullock et al. 2001; Zhao et al. 2009).  The bottom panel of Figure\ \ref{dependencies.fig} plots the same models presented above adopting a range of concentration parameters as colored lines.  The impact of changing $c$ on the normalized rotation curve is only marginal compared to the uncertainties throughout the region we probe with our stack.  Note that this behavior is partly, as in the case of adiabatic contraction, due to the normalization of our model in velocity.  {Increasing} $c$ at a given $m_{\rm{d}}$ does lead to a dark matter halo rotation profile with an overall higher rotation velocity, but this does not alter the shape in our models {significantly}.  At the lowest concentration explored ($c = 2$), our models indicate a steeper drop in the outer rotation curve as a signature of the decreasing imprint of dark matter on the (outer) potential, leading to a marginally better agreement with our stack.\\

\section{Discussion}

\subsection{An emerging picture of baryon dominance in high-z disks}
 {
{We have shown that the steep outer fall-off of our stacked \rc\ is consistent with non-zero, but remarkably low dark matter {contributions} to the typical rotation curves of high-z SFGs even at several effective radii.  Our findings confirm the previous results by {Genzel et al. (2017)}, who analyzed in detail six individual rotation curves at $0.85 < z < 2.4$, from which four are part of our stack.  Our stacked rotation curve and these individual rotation curves decline in a similar fashion at large radii.  Our results are also {in} excellent agreement with the findings of current studies {showing} that the dynamical mass budget of massive high-z SFGs is dominated by baryons within the disk scale (Burkert et al. 2016; Price et al. 2016; van Dokkum et al. 2015; Wuyts et al. 2016; {\"U}bler et al. 2017; Stott et al. 2016), and confirms that this finding holds even in the outer disk.}  Most importantly, our results are largely insensitive to the inferred mass-to-light conversions (affecting the inference of both stellar and gas masses), which represent a substantial uncertainty inherent to previous methods of inferring baryonic fractions at high redshift.  We demonstrate that our results are unlikely to be affected by the details of the inner baryonic mass distribution in galaxies.}  \\  

Extrapolating the inferred amount of baryons and dark matter on scales of the (outer) disk to the virial scale implies high disk mass fractions $m_{\rm{d}}$ within the dark matter halo. Our stacked rotation curve points to a baryonic $m_{\rm{d}}$ that is in rough agreement with our fiducial comparison value of $m_{\rm{d}} \sim 0.05$ within the $\sim 1-\sigma$ uncertainties (accounting for the pressure support of the outer disk), with this value being consistent with abundance matching results (Burkert et al. 2016).  Although still lacking strong significance, even higher baryonic fractions (up the cosmic baryon fraction of $m_{\rm{d}} = 0.17$, i.e. higher than anticipated by abundance matching) are formally preferred, again in agreement with the findings from {Genzel et al. (2017)}.\\
{
All these results paint a consistent picture: galaxies at high redshift are strongly baryon dominated. The redshift dependence implied by this picture has been observed, with Wuyts et al. (2016) showing that galaxies are more strongly baryon-dominated at {$z=2$ than $z=1$}.}  These findings also imply a highly efficient deposition and confinement of baryons available in a given DM halo to within the disk.  However, since {extrapolating} a value $m_{\rm{d}}$ from the disk to the virial scale strongly depends on the assumption of underlying dark matter density profiles, our results might indicate that the latter are less concentrated at high redshift than currently predicted by N-body simulations.  In this context, we have also shown that the scenario of adiabatic contraction at a given $m_{\rm{d}}$ considered in this work is only very weakly imprinted in the predicted models and thus cannot be tested with our stacked rotation curve.

\subsection{Implications for outer disk structure at high redshift}

Both the comparison of our stack with modeled rotation curves {and} the finding that the outer slope of our stack correlates with $V_{\rm{rot}}/\sigma_0$ suggest the presence of pressure gradients even in the outer disk at several effective radii. {{ For the models presented above that include the effect of pressure support we have assumed that $\sigma_0$ stays constant out to at least $\sim 2.4$ $R_{\rm{turn}}$ within a given galaxy.  As $\sigma_0$ can be measured for typical \textit{individual} SFGs {out to} $\sim R_{\rm{turn}}$, this assumption still remains to be confirmed.}}  Interestingly, a constant level of dispersion for a hydrostatic pressurized gas disk {implies} that the scale height $h_z$ increases exponentially with radius (Burkert et al. 2010):

\begin{equation}
h_z = \frac{\sigma_0^{2}}{\pi G \Sigma_0} exp\Big(\frac{r}{r_d}\Big),
\label{scale_height.eq}
\end{equation}

which can in principle be tested by observations.  However, such rapidly increasing scale heights of high-z disk galaxies in the outer regions have not been reported based on the currently available HST imaging, although surface brightness limitations severely hamper scale height measurements of faint outer disks. Observations of isolated disk galaxies in the local universe have shown that local spirals have radially constant scale heights in their {neutral gas and stellar} distribution (e.g. Kregel et al. 2004; van der Kruit \& Freeman 2011{, and references therein}). Consequently, the resulting velocity dispersion decreases with radius, leading to a weak imprint of pressure gradients in their outer disk (see also Dalcanton et al. 2010).  Assuming a constant scale-height throughout the disk in our models would imply $\sigma_0 = \sigma_0(r)$ decreasing exponentially with radius.  {In this limit there is little room for pressure support beyond the turn-over radius, in contradiction to the findings we present here.} \\

Another consequence of a constant $\sigma_0$ profile is a pressure-driven truncation of the disk where Equation\ \ref{pressure.eq} predicts $v_{\rm{rot}}(r)=0$.  These are expected at $R \gtrsim 2.4 R_{\rm{turn}}$, depending on $V_{\rm{rot}}/\sigma_0$ and $m_{\rm{d}}$ of a given galaxy. 
For comparison, photometric as well as kinematical signatures of truncations are frequently observed in local disk galaxies at galactocentric radii of  $3.5 - 4 r_d$ (e.g. van der Kruit \& Freeman 2011 and references therein).  Their origin, however, is believed to be connected to a minimum gas density threshold for star formation (Kennicutt 1998, Schaye et al. 2004), the current boundary of star formation propagating outwards within the disk (e.g. Larson 1976), or a maximum angular momentum of the protogalaxy (Fall \& Efstathiou 1980).  Such outer radial truncations might be present among high-z SFGs, with their observational signatures being expected at very faint surface brightness levels which strongly challenges even deep observations at high redshift.  Nelson et al. (2016) employ a stacking approach with the WFC3 grism spectroscopic data from the 3D-HST survey for $\sim 3000$ galaxies at $z \sim 1$ to derive extended radial profiles of \Ha\ and optical continuum emission.  The authors find that the 
resulting (outer) profiles {are in good agreement} with being exponential until $\sim 5 r_d$, where truncations in local disks are 
observed.  However, our findings motivate deep and sensitive future observations tracing both the rest-frame optical/NIR stellar light and tracers of the (ionized) gas-phase in detecting signatures of outer truncations as well as the vertical structure of disk galaxies at high redshift.

\section {Conclusions}

{In this work we have presented an analysis aimed at constraining the typical outer rotation profiles of disk galaxies at high redshift to shed further light on the structure and baryonic fractions of their outer disks.}  {We combined} resolved \Ha\ kinematic data for a sample of 101 massive {SFGs} at $0.6 \leqslant z \leqslant 2.6$ based on the large seeing limited \kd dataset with additional targets from the SINS/zC-SINF AO survey.  {We derived} a stacked rotation curve representative {of} the kinematics of {high-z SFGs in the mass range {$9.3 \lesssim \log{M_*/M_{\sun}} \lesssim 11.5$}. We} compared the resulting stack with modeled rotation curves to constrain the relative baryonic fractions and kinematic properties in the outer disk regions.  Our main results are the following:

\begin{itemize}

\item Through stacking we were able to constrain a representative rotation curve for our sample out to $\sim 2.4 R_{\rm{turn}}$ (corresponding to $\sim 4$ effective radii), allowing for a systematic view {into} outer disk kinematics at $z \sim 0.6 - 2.6$.  Our stacked rotation curve exhibits a decrease in rotation velocity beyond the turn-over radius down to { { $\sim 62 \%$} of the maximum normalized velocity $V_{\rm{max}}$}, confirming the drop seen in {six} individual galaxies ({Genzel et al. 2017}) as a representative feature for our sample of high-z disk galaxies.  The drop seen in our stacked rotation curve strikingly deviates from the average rotation curves of local spirals at the same mass at $> 3 \sigma$ significance level.

\item The comparison with modeled rotation curves shows that the falling stacked rotation curve can be explained by a high fraction of the {galaxy's total baryonic disk mass} relative to the dark matter halo ($m_{\rm{d}} \gtrsim 0.05$), in combination {with pressure support in the outer disk}.  These results are in good agreement with recent studies demonstrating that star-forming disks at high redshift are strongly baryon dominated at least in their inner parts (e.g. Burkert et al. 2016; Wuyts et al. 2016; Price et al. 2016; Stott et al. 2016; {Genzel et al. 2017}), with our findings being insensitive to light-to-mass conversions.

\item Splitting our sample into bins of $V_{\rm{rot}}/\sigma_0$, we show that the outer slope of our stacked rotation curves correlates with the amount of pressure support of the outer disk, confirming that pressure gradients {play an important role in explaining} the outer fall-off observed for our stacked rotation curve. 

\item We demonstrate that these results are largely independent of our model assumptions such as the absence or presence of a central stellar bulge, the possible effect of adiabatic contraction at fixed $m_{\rm{d}}$, and the halo concentration parameter.

\end{itemize}

\vspace{10mm}

\section*{}
This study is based on observations obtained at the Very Large Telescope of the European Southern Observatory, Paranal, Chile (ESO Programme IDs 075.A-0466, 076.A-0527, 079.A-0341, 080.A-0330, 080.A-0339, 080.A-0635, 081.B-0568, 081.A-0672, 082.A-0396, 183.A-0781, 087.A-0081, 088.A-0202, 088.A-0209, 091.A-0126, 092.A-0091, 093.A-0079, 094.A-0217, 095.A-0047, 096.A-0025, 097.A-0028,097.A-0353).\\
We thank the staff of Paranal Observatory for their support.  We are grateful to C.M. Carollo, A. Renzini, and S. Lilly for
useful suggestions and insightful comments on this work.  We thank S. Courteau, K. Freeman, O. Gerhard, M. Verheijen and R. Herrera-Camus for stimulating discussions.  J.C.C. Chan acknowledges the support of the Deutsche Zentrum fuer Luft- und Raumfahrt (DLR) via Project ID 50OR1513.  M.F. and D.W. acknowledge the support of the Deutsche Forschungs Gemeinschaft (DFG) via Projects WI 3871/1-1, WI3871/1-2.

\vspace{10mm}

\appendix
\setcounter{figure}{0}

\section {Detailed sample properties of stacked galaxies}
\label{A.app}

In the following, we will present the properties of the stacking sample selected as described in Section 2.3 and the underlying \kd + SINS/zC-SINF AO samples of detected and {spatially} resolved galaxies.  For this purpose, we consider the distributions in $M_*$, SFR, $R_{\rm{e}}$, and \sersic\ index.  Moreover, we determine the overlap of each galaxy with the main sequence by computing the logarithmic main sequence offset, $\Delta(\rm{MS})$, in specific SFR (sSFR) given by 

\begin{equation}
  \Delta(\textrm{MS}) = \log(\textrm{sSFR/sSFR}_{\rm{MS}}(z,M_*)),
\label{delta_ms.eq}
\end{equation}

where $\textrm{sSFR}_{\rm{MS}}(z,M_*)$ is the expected sSFR on the main sequence given a galaxy's redshift $z$ and stellar mass $M_*$.  In order to compute $\textrm{sSFR}_{\rm{MS}}(z,M_*)$, we adopt Equations 5 and 6 from Wisnioski et al. (2015), who use the parametrization of the main sequence as measured by Whitaker et al. (2014).\\

We also compute the offset of each target from the average $R_{\rm{e}} - M_*$ relation for late-type galaxies derived from van der Wel et al. (2014a).  {This offset is computed by deriving the { $R_{\rm{e}} - M_*$ relation at the redshift of each target, taking into account the redshift evolution of both zero-point and slope of the $R_{\rm{e}} - M_*$ relation { for late-type galaxies by van der Wel et al. (2014a)}}. We also take into account} the conversion of $R_{\rm{e}}$ derived from $H$-band to a rest-frame wavelength of 5000 \AA{} as done in van der Wel et al. (2014a).  The latter yields a size correction of $\sim 11 \%$ on average for our stacking sample.

The properties of both stacking sample and the \kd + SINS/zC-SINF AO samples of detected and resolved galaxies are {listed} in {Table A1,  which gives the} ranges, mean, and median values of their distribution in $M_*$, SFRs, $R_{\rm{e}}$, $\Delta(\rm{MS})$, and \sersic\ index $n$.  The respective distributions are displayed in Figure A1.  {To evaluate how representative our sample is compared to the underlying population of SFGs, we also plot the distribution of star-forming systems within the parent 3D-HST sample in the same range {(i.e. with $sSFR > 0.7/t_{\rm{Hubble}}(z)$, $\log{M_*/M_{\sun}} > 9.3$}, {rejecting galaxies that only have a photometric redshift}).}

To further facilitate a statistical comparison between properties of the stacking sample and the \kd + SINS/zC-SINF AO samples, we perform two-sided Kolmogorov - Smirnov (K-S) tests on the distributions of the above mentioned parameters.  The resulting probability values (`p-values', $p$) are indicated for each panel in Figure A1.\\

\begin{table*}[t]
\caption[]{Properties of the stacking sample and the \kd + SINS/zC-SINF AO samples of detected and resolved galaxies.  Stellar masses $M_*$, SFRs and intrinsic effective Radii $R_{\rm{e}}$ are given in units of $M_{\sun}$, $M_{\sun}/ yr$ and in kpc, respectively.}
{\small
\centering
\begin{tabular}{lllccccccc}
\multicolumn{1}{c}{}                     & \multicolumn{1}{c}{} &  & \multicolumn{3}{c}{Stacking sample}     &  \hspace{2.0mm} & \multicolumn{3}{c}{\kd + SINS/zC-SINF AO}  \\
Property               			 &                      &  & Range        & Median  & Mean &  & Range         & Median & Mean \\ \hline
\multirow{2}{*}{$\log(M_*)$}             & z \textless 1.2      &  & [9.25,11.23] & 10.61   & 10.50     &  & [9.20,11.23] & 10.42    & 10.34      \\
\vspace{01.5mm}                          & z \textgreater 1.2   &  & [9.66,11.46] & 10.61   & 10.63     &  & [9.36,11.50] & 10.43    & 10.49      \\
\multirow{2}{*}{$\log(\rm{SFR})$}             & z \textless 1.2      &  & [0.36,2.09]  & 1.39    & 1.45      &  & [-1.05,2.11] & 1.22    & 1.34      \\
\vspace{01.5mm}                          & z \textgreater 1.2   &  & [1.18,2.55]  & 1.91    & 2.02      &  & [-0.65,3.08] & 1.78    & 2.04      \\
\multirow{2}{*}{$R_{\rm{e}}$}            & z \textless 1.2      &  & [2.32,8.58]  & 4.61    & 4.95      &  & [0.53,10.22] & 3.85    & 4.44      \\
\vspace{01.5mm}                          & z \textgreater 1.2   &  & [1.40,9.30]  & 4.49    & 4.63      &  & [0.25,9.54]  &  3.19    & 3.44     \\
\multirow{2}{*}{$\Delta(\rm{MS})$}       & z \textless 1.2      &  & [-0.50,0.93] & 0.20    & 0.15      &  & [-2.27,1.49] & 0.06    & 0.01      \\
\vspace{01.5mm}                          & z \textgreater 1.2   &  & [-0.62,0.77] & 0.05    & 0.03      &  & [-2.93,1.45] & 0.03    & 0.01      \\

\multirow{2}{*}{$\log(R_{\rm{e}}/R_{\rm{e,vdW14}}(M_*,z))$}       & z \textless 1.2      &  & [-0.29,0.81] & 0.07    & 0.06      &  & [-0.87,0.81] &  0.01    & -0.01      \\
\vspace{01.5mm}                                                   & z \textgreater 1.2   &  & [-0.45,0.38] & 0.08    & 0.07      &  & [-1.11,0.63] & -0.01    & 0.06      \\

\multirow{2}{*}{$n$}                     & z \textless 1.2      &  & [0.5,4.2]    & 1.31    & 1.64      &  & [0.2,8.0]$^{*}$ & 1.43    & 1.84      \\
\vspace{01.5mm}                          & z \textgreater 1.2   &  & [0.2,7.9]$^{*}$    & 0.68    & 1.28      &  & [0.2,8.0]$^{*}$ & 1.11    & 1.59      \\ \hline
\end{tabular}
\\\hspace{\textwidth}
\tablenotetext{*}{These boundaries are set within the \sersic\ profile fitting; for details see Lang et al. (2014) and van der Wel et al. (2012).}
\label{Table_values.tbl}}
 \vspace{3mm}
\end{table*}

 \begin{figure*}[t]
 \centering
   \includegraphics[width=0.85\textwidth]{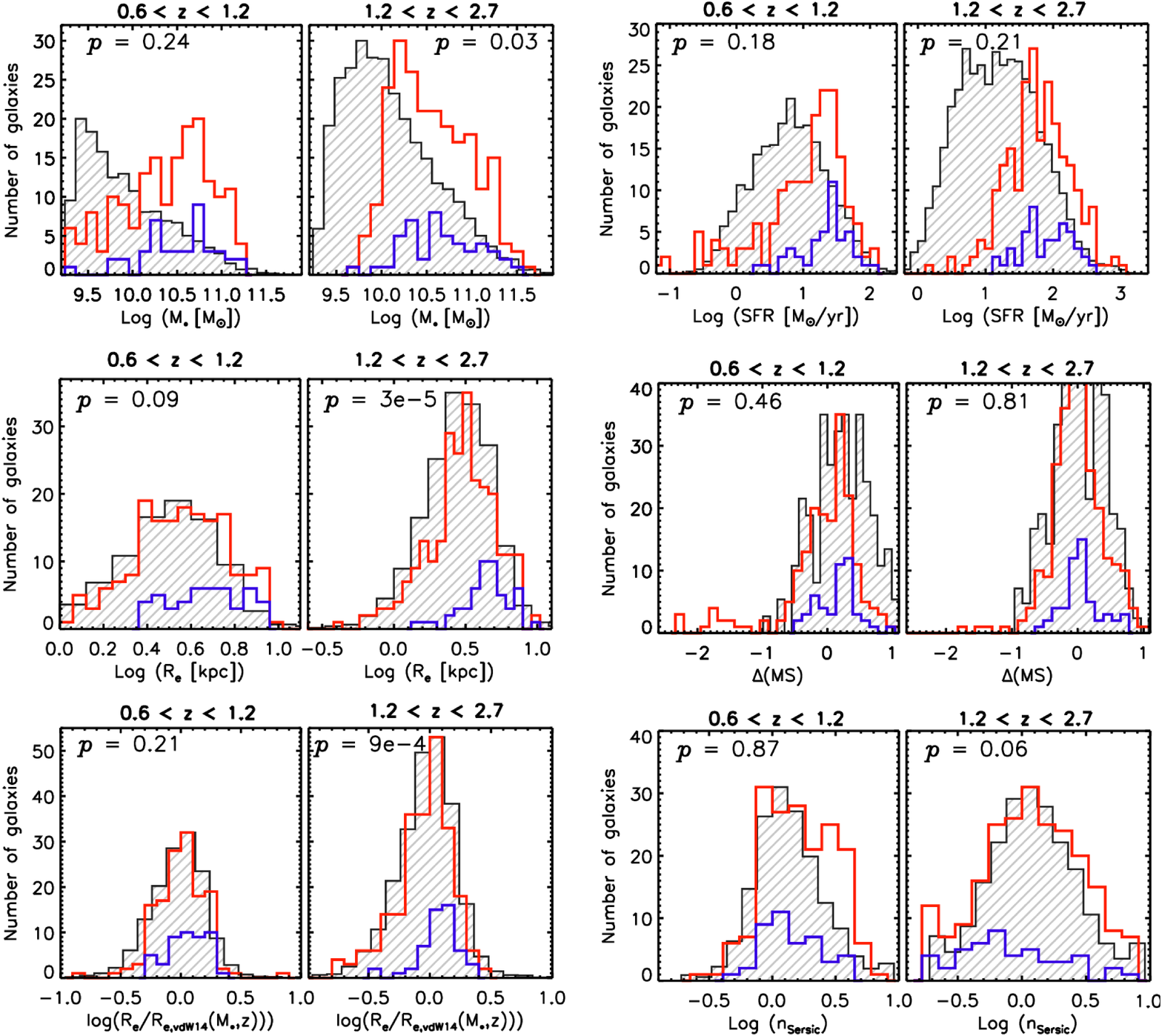}
 \caption[Distributions of galaxy parameters]{Distributions of {stellar }masses $M_*$, SFRs, intrinsic effective {radii} $R_{\rm{e}}$, main sequence offsets $\Delta(\rm{MS})$, offsets from the $R_{\rm{e}} - M_*$ relation, and \sersic\ indices $n$ of the stacking sample (blue) and the \kd + SINS/zC-SINF AO samples of detected and resolved galaxies (red).  The distributions are shown in two redshift bins. The p-value from the K-S test on both respective distributions are furthermore indicated in each panel. {The underlying 3D-HST population of SFGs above $\log{M_*/M_{\sun}} > 9.3$ is shown as grey histograms, normalized in peak number to the red histograms.}}
 \label{Histogram_values.fig}
 \vspace{3mm}
 \end{figure*}

Figure A1, as well as the values in Table A1 demonstrate that the stacking sample yields an overall fair representation of the star-forming main sequence population within the full redshift range probed. Although we significantly reduce the number of galaxies when selecting the sample of stacked galaxies, we find only little bias in the average galaxy properties {considered} here between the stacking sample and the \kd + SINS/zC-SINF AO samples of detected and resolved galaxies. Most notably, the stacking sample shows mild offset towards more massive systems and larger sizes, with the former being only marginally significant.  {We conclude the following:}

\vspace{4mm}
\begin{itemize}

\item The stellar mass distribution of the stacking sample {in terms of} its range and average shows good overlap with the \kd + SINS/zC-SINF AO samples, albeit with a shift towards more massive systems. While the latter extend to lower mass systems below $\log{M_*} < 10$, this mass range scarcely overlaps with our stacking sample.  This is connected to the fact that this lower-mass regime contains on average both smaller systems and a smaller fraction of rotation-dominated galaxies (see Wisnioski et al. 2015).  Both of these effects cause galaxies to be preferentially rejected from our sample ({see below for a discussion about sizes}).  {By  design of the surveys, both kinematic \kd + SINS/zC-SINF AO samples and hence our stacking sample features a flat mass distribution with the emphasis on more massive systems, compared to steeply dropping mass function of the mass-complete 3D-HST parent population of SFGs.}

\item Considering the coverage in SFR, both {the stacking and the} \kd + SINS/zC-SINF AO samples have very similar average values probed for both redshift ranges.  The ranges in SFR probed by the stacking sample are slightly smaller, since in particular the SFRs do not extend to the low values well below the MS, especially at $z < 1.2$, which is, however, not reflected by a statistically significant difference.  {Since the \kd + SINS/zC-SINF AO samples are weighted towards more massive systems compared to the underlying population of SFGs, their SFR distribution is offset towards higher values. However, as shown by investigating the main sequence offset discussed below, the SFRs of our kinematic samples overlap well with those of the underlying population at a given stellar mass.}

\item Our stacking sample overlaps well with the main sequence in both redshift ranges as indicated by the average main sequence offset  $\Delta(\rm{MS})$.  {The numbers given in Table A1} point out that the \kd + SINS/zC-SINF AO samples extend to somewhat lower sSFRs into the regime well below the MS, compared to our stacking sample.  Given that our selection requires tracing the \Ha\ kinematics at least out to radii where the rotation curve flattens, we thus are not able to include targets with low \Ha\ surface brightness well below the MS. 

\item Turning to the size distributions, {our stacking sample overlaps well} with the underlying size distribution of late-type galaxies at redshifts $z < 1.2$ as indicated by the right upper panel of Figure\ \ref{sample_properties1.fig}.  However, we find that the average size of targets in our stacking sample is slightly larger than for the \kd + SINS/zC-SINF AO samples, most importantly at $z > 1.2$.  Accounting for the $R_{\rm{e}}$  evolution with redshift, the bias towards large sizes is still reflected in a positive offset with respect to the $R_{\rm{e}} - M_*$ relation at $z > 1.2$.  This offset is at most $\log(R_{\rm{e}}/R_{\rm{e,vdW14}}(M_*,z)) \sim 0.08$, thus small compared to logarithmic scatter of $\sim 0.16 - 0.19$ in the  $R_{\rm{e}} - M_*$ relation.  This remaining bias in size is qualitatively consistent with our mock analysis presented in Appendix C.

\item The distribution of \sersic\ indices as measured on the rest-frame optical light distributions for our stacking sample and for the \kd + SINS/zC-SINF AO samples are very similar and in agreement with being exponential.
 \end{itemize}

\section {Using \Re\ to calibrate $R_{\rm{turn}}$}
\label{B.app}
\setcounter{figure}{0}

\begin {figure*}[t]
\centering
 \includegraphics[width=0.41\textwidth]{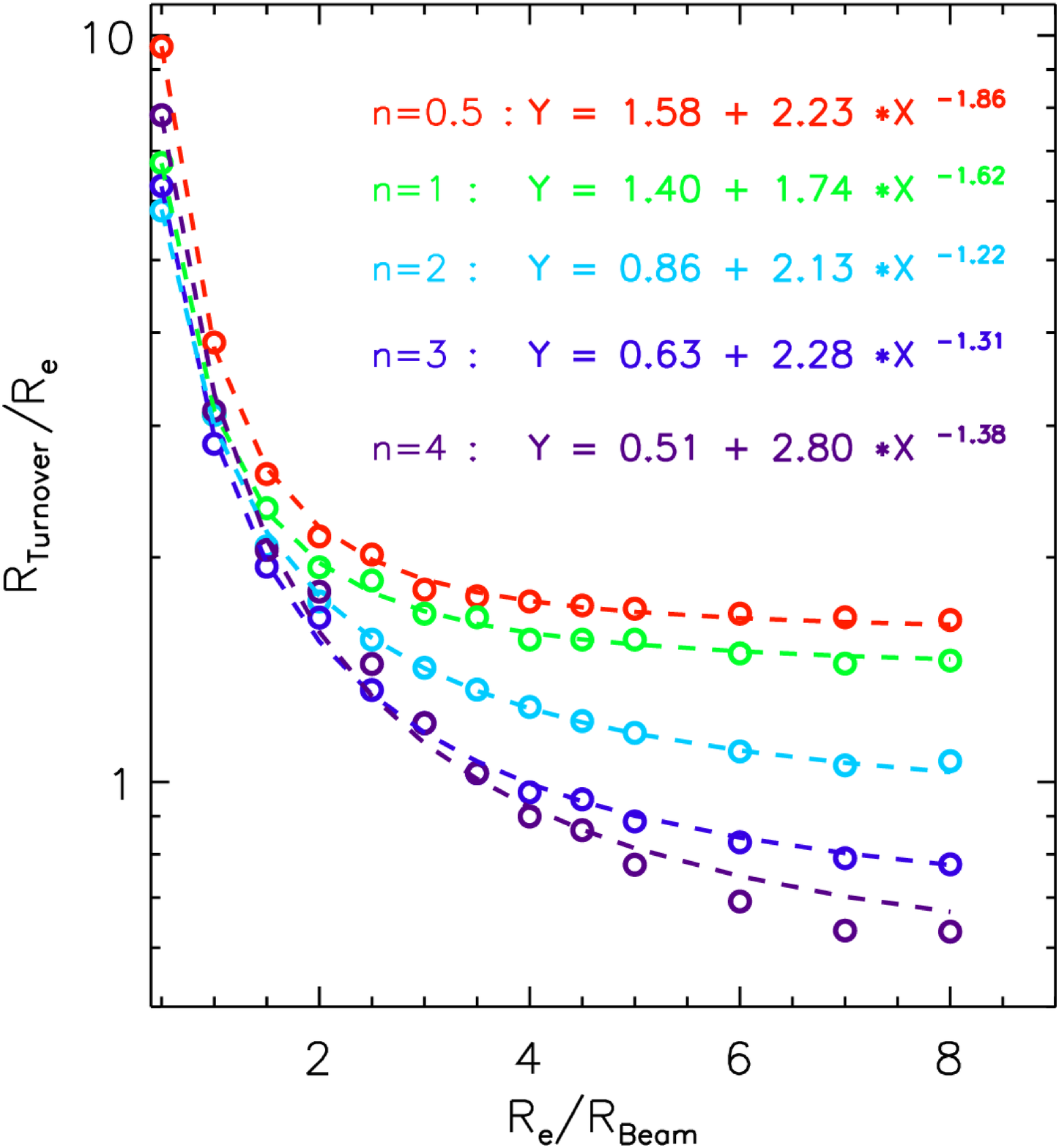}
 \includegraphics[width=0.44\textwidth]{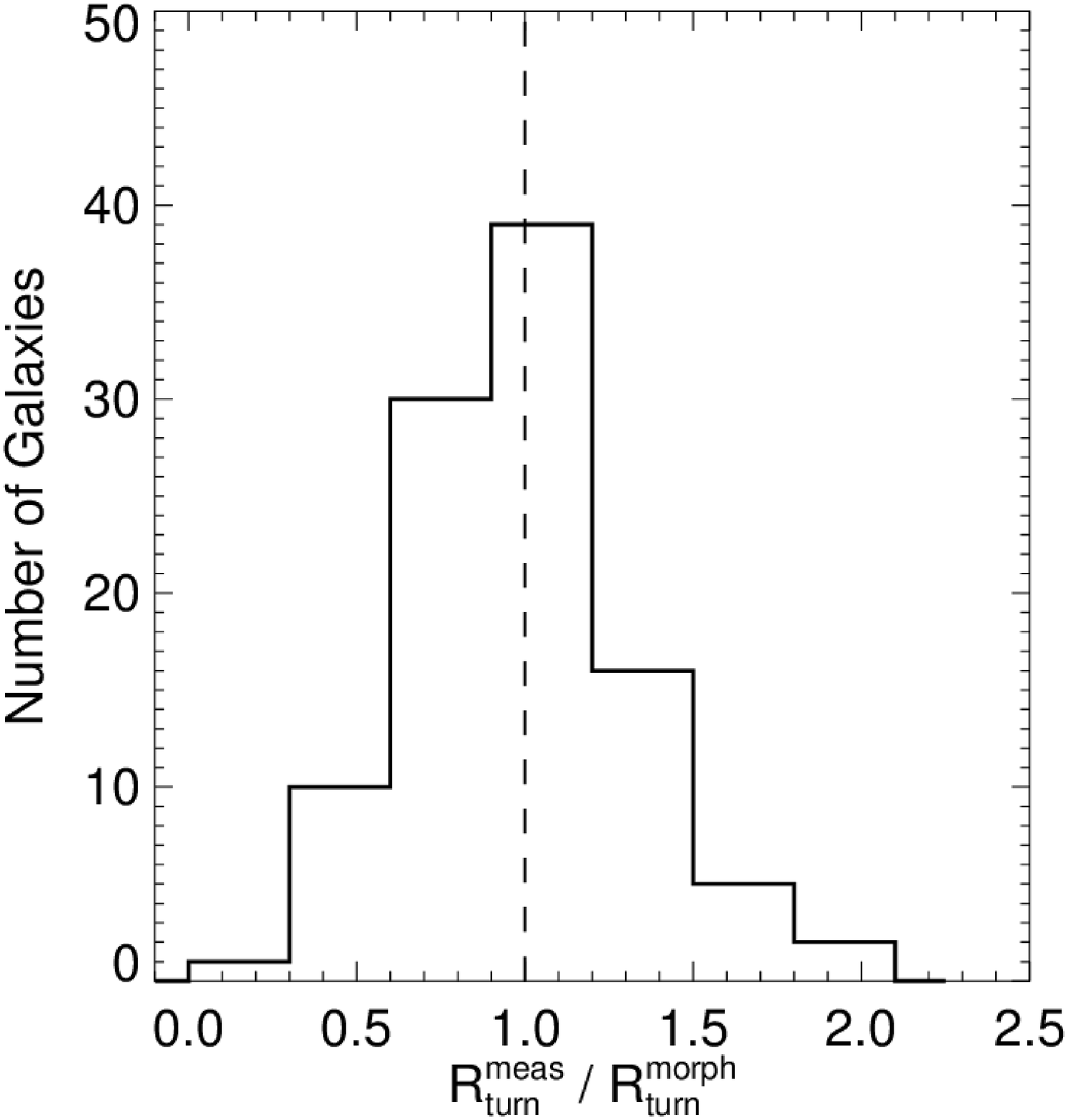}
\includegraphics[width=0.53\textwidth]{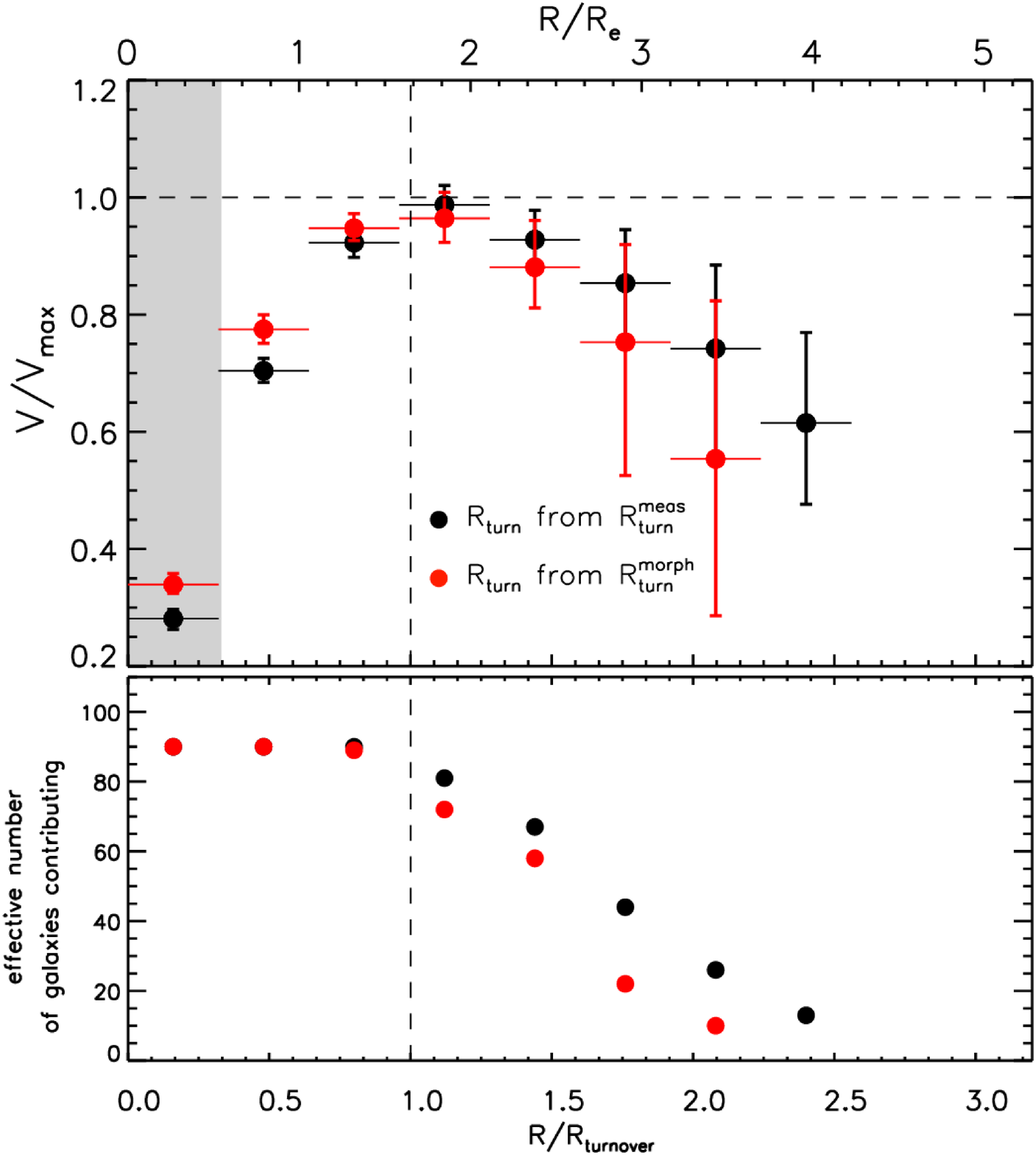}
\caption[Modeled RCs with varying \sersic\ index]{Top left: The ratio between the \rturnm\ and intrinsic effective radius $R_{\rm{e}}$ as a function of $R_{\rm{e}}/R_{\rm{beam}}$ derived from modeled mock galaxies.  The points are color-coded by \sersic\ index.  The fits are shown as dashed lines, and their parameterizations are also shown. Top right: Distribution of calibrated versus observed turn-over radii of the stacked galaxy sample.  The dashed line indicates a ratio of 1. Bottom: Stacked rotation curve shown in Figure\ \ref{rc.fig} (black symbols) together with an alternative stacked rotation curve derived using \rturnm\ derived from $R_{\rm{e}}$ (red symbols).  The shaded area marks the half-light beam size of the average PSF observed for our sample. 
}
\label{simulations2.fig}
 \vspace{3mm}
\end {figure*}

As discussed in Section\ \ref{normalization.sec}, the measurements of \rturno\ are based on the assumption of exponential mass distributions and are possibly affected by central stellar bulges.  To validate our measured \rturno, we employ an alternative approach to derive \rturn\ independently of the parametrization and fit of the observed rotation curve, taking into account deviations from exponential profiles.  

Thus we aim to convert a measurement of intrinsic baryonic size, $R_{\rm{e}}$, provided by the \sersic\ profile parametrization of the rest-frame optical morphology, into an observed turn-over radius, using information on the \sersic\ index $n$ and inclination (given by the measured axial ratio $b/a$).  The resulting turn-over radii are referred to as `\rturnm', in contrast to the turn-over radii determined from the individual \rcs\ (\rturno).  

In order to derive a conversion between a given \rturnm\ and the \Re\ that can be applied to our sample, we use rotation curves of simulated galaxies.  We construct noise-free mock data cubes using the DYSMAL code (Davies et al. 2011, see also Burkert et al. 2016) with a range of \sersic\ indices, sizes, and inclinations.  The DYSMAL code creates mock data cubes given intrinsic galaxy parameters, which are convolved spectrally and spatially with an instrumental beam to mimic real 
observations.  From the data cubes, rotation curves are then extracted along an artificial slit similar to the methodology performed on our data.  {In} the modeling, we consider a finite thickness of the disk with $q = 0.2$, using the recipe in Noordermeer (2008) to calculate the corresponding rotation curves in DYSMAL, {neglecting the effect of pressure support.}  For each model \rc, we measure \rturnm\ as the peak position of the rotation curve to compute the ratio \ret.\\

Due to the difference in rest-frame wavelength sampled by $H$-band observations over the range  $z \sim 0.6 - 2.6$ ($\sim 8000$ \AA{} at $z\sim 1$, $\sim 5300$ \AA{} at $z\sim 2$), color gradients within galaxies might affect our size measurements.  To mitigate this effect we convert our size measurements to a rest-frame wavelength of 5000 \AA{} by applying a redshift and mass dependent correction from van der Wel et al. (2014a).   

\ret\ shows a dependence on the intrinsic galaxy size compared to the size of the beam ($R_{\rm{e}}/R_{\rm{beam}}$) and the \sersic\ index, but very little on inclination.  The upper left panel of Figure B1 illustrates these dependencies by showing the ratio \ret\ as a function of $R_{\rm{e}}/R_{\rm{beam}}$, plotted for different adopted \sersic\ indices in different colors.  As expected, \ret\ is strongly anti-correlated with  $R_{\rm{e}}/R_{\rm{beam}}$ since more severe beam smearing will shift the peak velocity of a rotation curve out to larger radii.  
We derive conversion factors based on our grid of models, which are then applied on a galaxy-to-galaxy basis, depending on $R_{\rm{e}}/R_{\rm{beam}}$, \sersic\ index, and inclination.  The ratio between \rturnm\ and \rturno\ for our sample galaxies is displayed in the upper right panel of Figure B1.  Overall, the data show a very good correspondence between \rturnm\ and \rturno\, with a median ratio between \rturno\ and \rturnm\ of $\sim 0.98$.  The scatter in the distribution is $\sim 0.3$ and stems likely from uncertainties in the determination of $n$, $b/a$, and $R_{\rm{e}}$ from fits to the $H$-band light profiles as well as uncertainties in the measurement of \rturno.\\

Following our stacking methodology with the \rturnm\ values adopted, we derive a stacked rotation curve shown in the bottom panel of Figure B1 (red symbols), which is compared to our fiducial \rturno\ - normalized stack in black.  As apparent in Figure B1, both stacks show a turn-over at the same location, which is expected from the good overall agreement between \rturno\ and \rturnm.  However, the distribution of \ret\ exhibits considerable scatter, which is potentially imprinted as larger uncertainties in the \rturnm\ - normalized stacked rotation curve.  Also, the number of galaxies as a function of radius is lower at $r > 1.5 R_{\rm{turn}}$, which likely contributes to the larger uncertainties of outer bins of the velocity curve. The outermost bin at $r = 2.4 R_{\rm{turn}}$ is not plotted for the \rturnm\ - normalized stack since only few ($< 10$) galaxies contribute, hampering a representative velocity measurement at this radius. 
Despite the larger uncertainties of the \rturnm\ - normalized stack, we measure an outer slope of $\Delta V / \Delta R = -0.32^{+0.18}_{-0.16}\, [V/V_{\rm{max}}$,$R/R_{\rm{turn}}]$, consistent with the outer slope of our nominal stack in Figure\ \ref{rc.fig} within the uncertainties. 
{For the above normalization in observed velocity for each target, we have used the values of $V_{max}$ as derived from the exponential fits discussed in Section 3.1 (i.e. $V_{max} = V($\rturno$)$).  When, instead, using the velocity observed at the predicted turnover \rturnm\, such that  $V_{max} = V($\rturnm$)$, we find a consistent slope of $\Delta V / \Delta R = -0.26^{+0.15}_{-0.14}\, [V/V_{\rm{max}}$,$R/R_{\rm{turn}}]$.}\\

{
With the \rturnm\ estimates in hand, we furthermore compare in Figure B2 the maximum radius out to which individual rotation curves can be measured, $R_{\rm{obs}}$, versus both the effective radius and expected turnover radius \rturnm\ .  As apparent in the left panel of Figure B2, the stacking sample has overall both larger $R_{\rm{e}}$ and $R_{\rm{obs}}$ compared to all disks in the  \kd + SINS/zC-SINF AO samples (i.e. targets which satisfy Criterion 1, but not 2 of our selection described in Section 2.3).  In addition, at fixed galaxy size, the stacking sample contains on average galaxies where rotation curves are available out to larger radii, indicating that $R_{\rm{obs}}$ (and thus $S/N$) is mostly responsible for determining whether a galaxy is selected for stacking.   Most of the galaxies in the stacking sample have rotation curves out to radii reaching or exceeding the expected turnover radius, whereas targets only satisfying Criterion 1 extend to lower radii and on average do not reach the expected turnover.  After accounting for the galaxies that have been rejected due to strong OH residual contamination according to Criterion 3, we find a minor subset of 32 galaxies with $R_{\rm{obs}}$/\rturnm$> 1$ that do not satisfy Criterion 2. \\
As this subset might represent galaxies with rotation curves that are rising in their outer parts, we add these additional targets to the stack presented in Figure B1, adopting a normalization radius of \rturnm\ and velocity of { $V_{max} = V($\rturnm$)$} for each target.  We find a resulting stack with a slope of $\Delta V / \Delta R = -0.33^{+0.15}_{-0.17}\, [V/V_{\rm{max}}$,$R/R_{\rm{turn}}]$, consistent with our stack presented in Figure B1 and our fiducial stack shown in Figure\ \ref{rc.fig}, and therefore conclude that those additional targets have average falling rotation curves in agreement with the stacked sample of 101 galaxies.  However, there is a notable difference in slope when comparing the new stack of 133 galaxies with our fiducial containing 101 targets using the same normalization methodology (i.e. both adopting \rturnm and { $V_{max} = V($\rturnm$)$}). 
We attribute this difference to the sample properties of the newly added 32 galaxies, in particular to their lower average $V_{\rm{rot}}/\sigma_0$ ratios (causing a decrease of $<V_{\rm{rot}}/\sigma_0 >$ from $\sim 6.3$ to  $\sim 5.8$), leading to a steeper outer fall-off according to Equation 8.}

\begin {figure}[h]
\centering

 \includegraphics[width=0.48\textwidth]{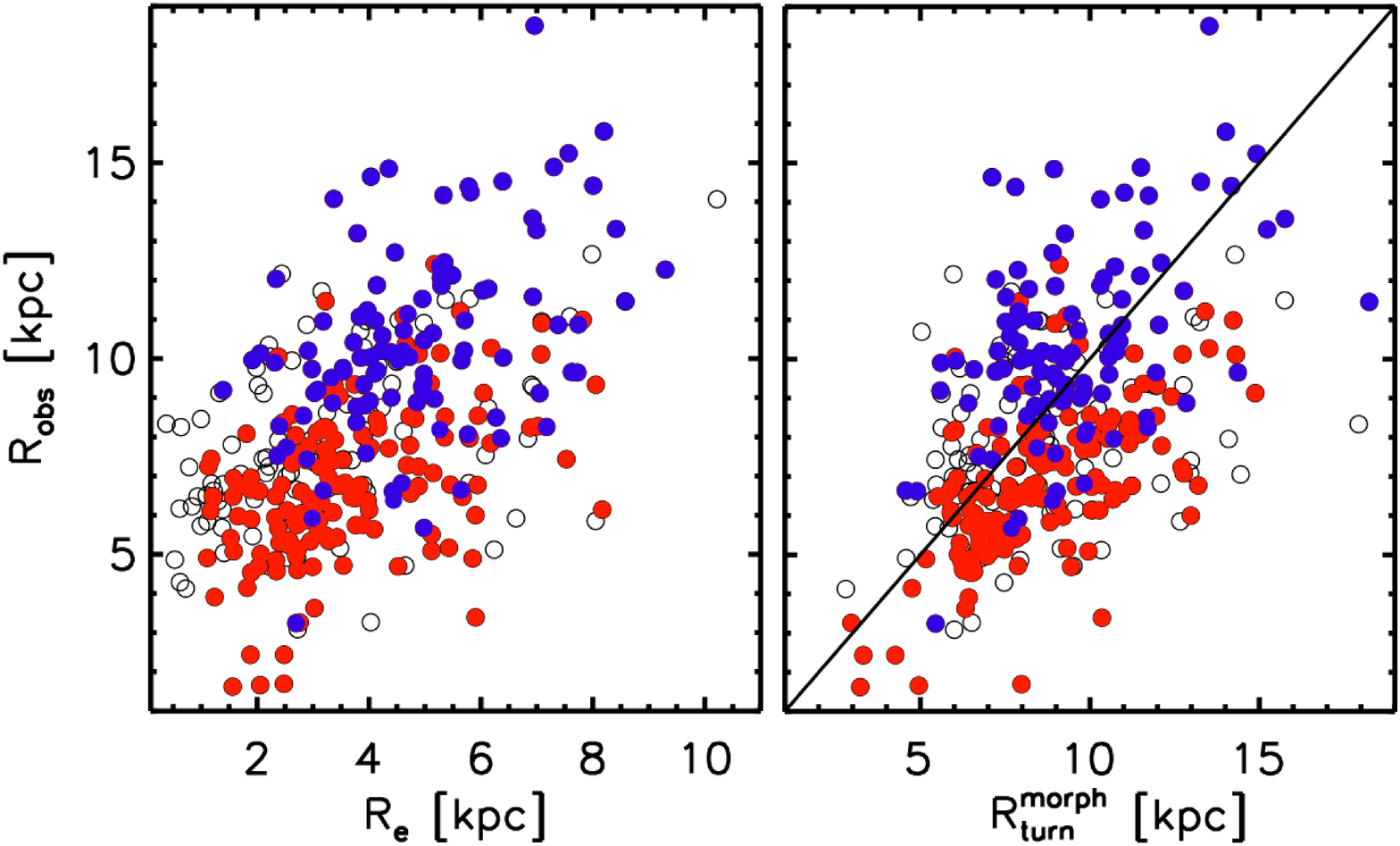}
\caption[Robs]{{Maximum radius at which individual rotation curves can be measured ($R_{\rm{obs}}$) versus the galaxy effective radius (left panel) and the expected turnover radius \rturnm\ (right panel).  Shown are all detected and spatially resolved \kd and SINS/zC-SINF AO targets (empty circles), with the galaxies satisfying only our selection criterion 1 above and the stacking sample overplotted as red and blue circles, respectively. Galaxies which have been rejected due to strong OH residual contamination are omitted.}}
\label{robs.fig} 
 \vspace{3mm}
\end {figure}

\section{Potential biases due to sample selection and normalization}
\label{Mock.sec}
\setcounter{figure}{0}

Here, we explore how the selection of galaxies for the stacking sample as well as the normalization of individual RCs with an exponential disk model could potentially affect the results on our stacked rotation curve.  In summary, we address the following two questions:

\begin{enumerate}

\item Does the selection of galaxies for stacking according to Criterion 2 (discussed in Section\ \ref{stacking_sample.sec}) bias our stacking sample towards galaxies with outer falling rotation curves as might be expected for DM-poor galaxies?

\item Under the assumption that the average outer rotation curves of massive SFGs at high redshift were \textit{rising} as expected for DM-dominated galaxies, could the resulting \textit{positive} outer slope be recovered by our stacking technique?
\end{enumerate}

\begin {figure*}[t]
\centering

 \includegraphics[width=0.75\textwidth]{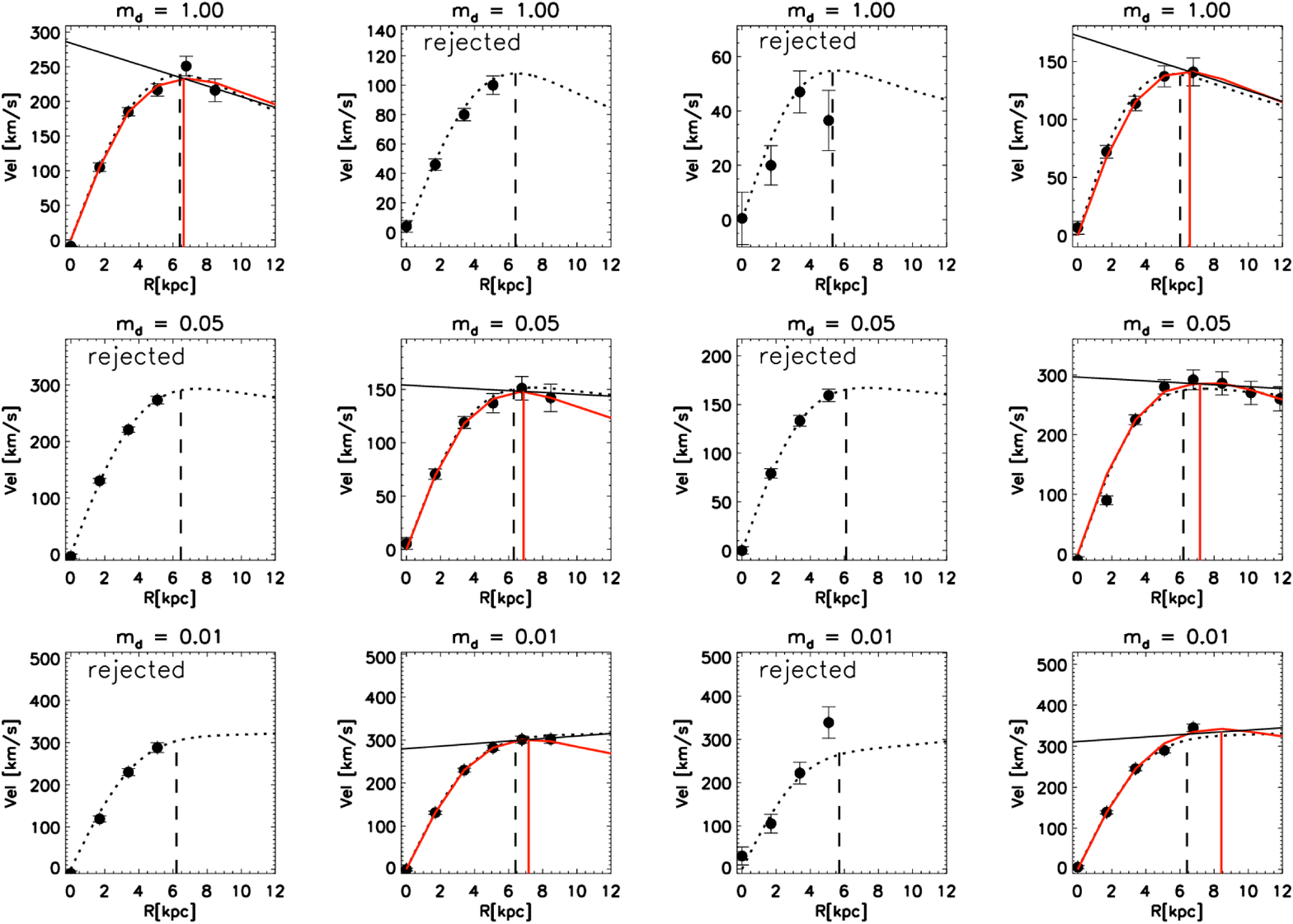}
\caption[Examples of simulated Mock rotation curves]{Simulated observed-frame and corresponding noise-free mock rotation curves, shown as symbols and dashed lines, respectively.  The different rows show models with decreasing $m_{\rm{d}}$ from 1 to 0.01. The four columns present different observed-frame realizations of the same rotation curve. The adopted $R_{\rm{e}}$ for all models shown is 3 kpc. {For each simulated model, \rturnm\ (i.e the expected peak of the observed-frame rotation curve without dark matter contribution) is indicated by the black dashed line.  In case the realization did not get rejected, fitted exponential disk models are overplotted in red, together with \rturno\ indicated as vertical red line. Additionally, the outer slope that was determined for the corresponding grid point [$R_{\rm{e}}$,$m_{\rm{d}}$] is {plotted} as solid black line.}}

\label{Mock_models.fig}
 \vspace{3mm}
\end {figure*}

Question 1 is important since Criterion 2 discussed in Section\ \ref{stacking_sample.sec} removes a large portion of targets from the available pool of rotating disks in the \kd and SINS/zC-SINF datasets.  In particular, we require individual RCs to show a sufficient change of slope, which is indeed expected for baryonic disks, and still so for DM-poor galaxies. However, outer RCs of DM-\textit{dominated} galaxies with outer rising profiles may exhibit much less outer flattening and therefore may be rejected more frequently, leading to a potential bias towards low dark matter fractions in our stacking sample.  We already demonstrated that the stacking sample has little to no bias in the average galaxy properties such as stellar mass, SFR, $\Delta(\rm{MS})$, and \sersic\ index compared to the \kd + SINS/zC-SINF AO samples { and only a small bias towards larger $R_{\rm{e}}$}.  However, there might exist additional galaxy properties connected to the dark matter fraction or the outer slope of RCs that are not examined here.\\
We raise Question 2 since the pure exponential disk model (without the contribution of dark matter) used to normalize our individual RCs provides a good description for \textit{inner} rotation curves until $R_{\rm{turn}}$, but does not reflect the outer \textit{rising} RCs for galaxies with significant DM components and thus potentially fails to recover rising outer slopes in our normalized stack.\\

{We address these questions by simulating 1D rotation curves. These simulated rotation curves} include a rotating exponential baryonic disk and {a} dark matter NFW halo for a range of sizes, masses and dark matter fractions.  The simulated rotation curves are derived {analogously to} the method presented in Section\ \ref{models.sec}, using a baryonic exponential disk with finite thickness, and NFW halo models.  For the combined disk+halo mock-RCs, we vary the effective size of the disk, $R_{\rm{e}}$, and the disk-to-dark matter halo mass fraction inside the halo, $m_{\rm{d}}$.  Pressure support and adiabatic contraction are neglected.  Mock-RCs are created for a grid of $R_{\rm{e}}$ ranging from 2 to 5 kpc, and for $m_{\rm{d}}$ in the range of 1 (no dark matter) to 0.01 (strongly dark matter dominated), adopting a total dynamical mass of $\log{M} = 11$.\\
At each given grid point [$R_{\rm{e}}$,$m_{\rm{d}}$] we create 25 observed-frame realizations of the rotation curve by applying a random inclination (i.e. scaling the RC by $\sin{i}$) and perturbation by typical uncertainties in velocity.  Furthermore, each realization is cut at a maximum galactocentric radius, $R_{\rm{obs}}$, simulating that individual observed RCs are traceable to a limiting radius.  Since we found that $R_{\rm{obs}}$ is, among various galaxy parameters, most {notably} correlated with $R_e$ for detected and resolved galaxies within the \kd dataset, we implement this dependence in our modeling.  The \sersic\ index $n$ of the disk is furthermore varied for each realization.  The values of $R_{\rm{obs}}$, $n$, and the typical uncertainty in velocity for each realization are determined by randomly drawing from Gaussian distributions.  The mean and scatter of those distributions are determined from the observed sample of detected and resolved \kd and SINS/zC-SINF galaxies.  
Then, each observed-frame rotation curve realization is convolved with a 1D Gaussian PSF of FWHM typical for our dataset, and the RC is sampled according to the pixel scale provided by the observed datacubes.

\begin {figure*}[tb]
\centering
 \includegraphics[width=0.45\textwidth]{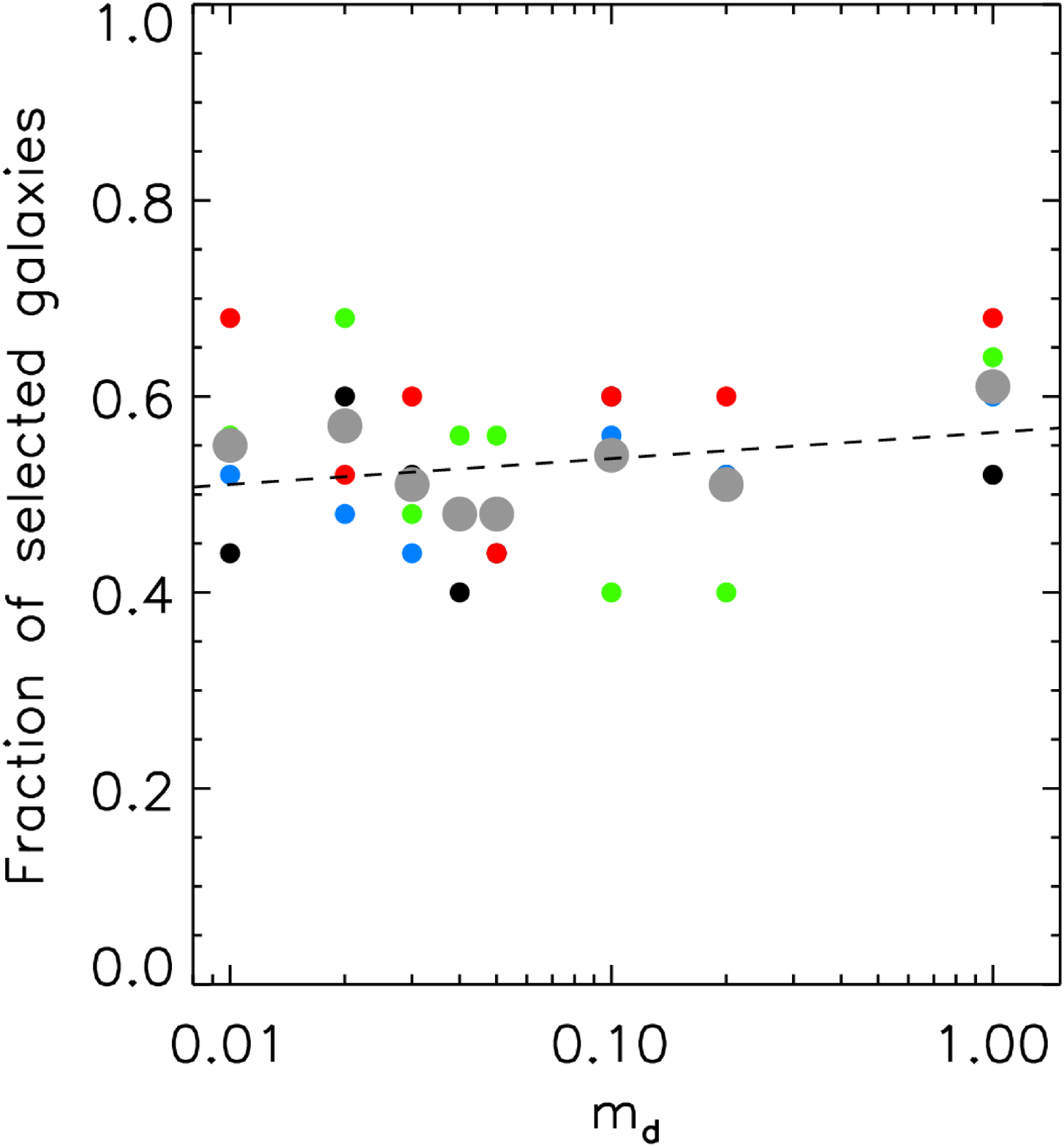}
 \includegraphics[width=0.48\textwidth]{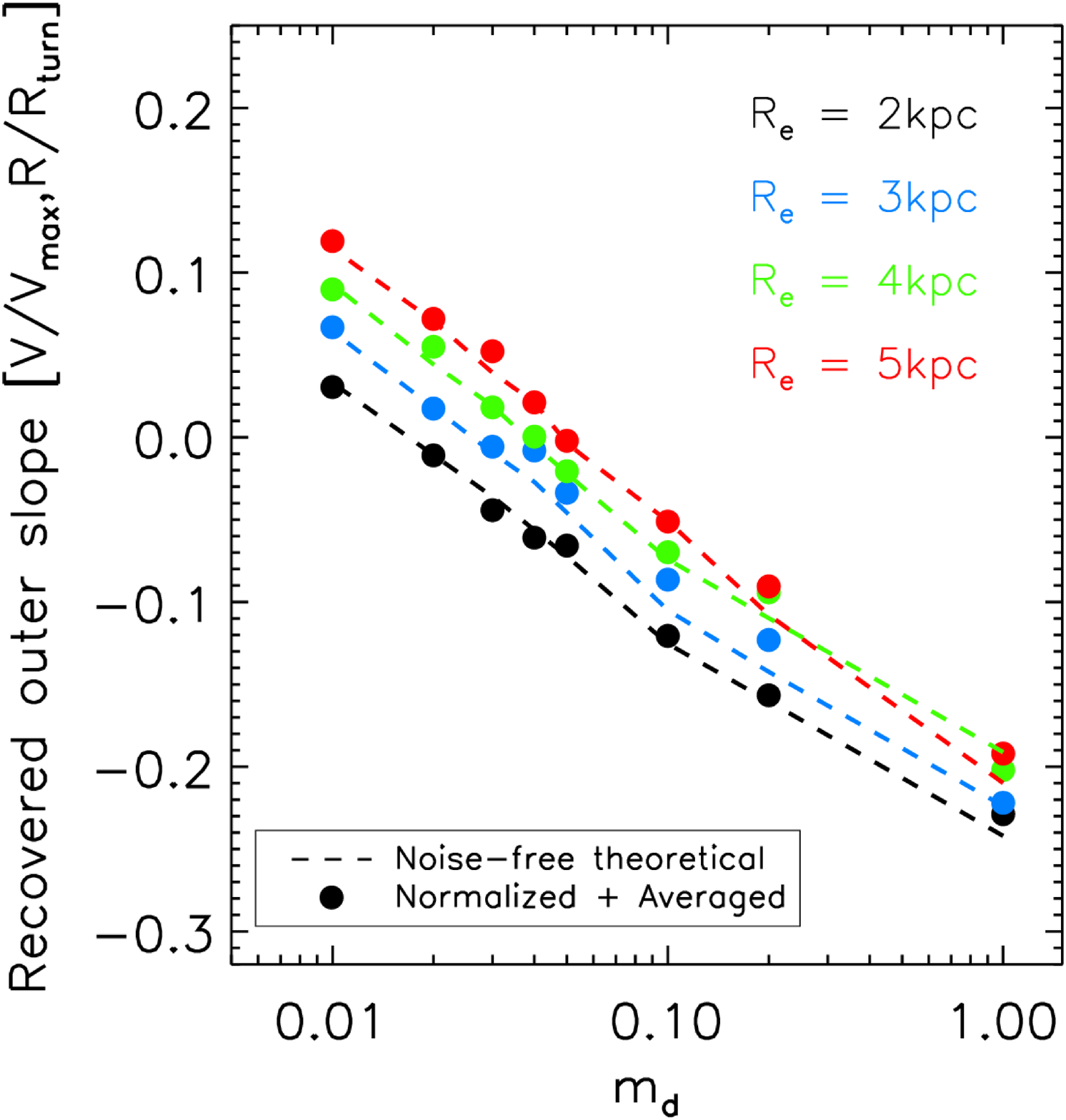}
\caption[Selection fractions and recovered outer slopes of mock rotation curves]{Left: Fraction of non-rejected mock RCs selected for stacking as a function of $m_{\rm{d}}$.  The colored symbols plot values for different adopted $R_{\rm{e}}$, while the grey circles represent the average over all $R_{\rm{e}}$ for a given $m_{\rm{d}}$.  The dashed line indicates the linear fit to all grey circles.  Right: Recovered outer slopes obtained from the noise-free RCs (dashed lines) as well as from the selected, normalized, and averaged mock-RCs (colored symbols).  The slopes are shown in units of our normalized coordinate frame $[V/V_{\rm{max}}$,$R/R_{\rm{turn}}]$, in analogy to the results presented in Section\ \ref{structure.sec}. }
\label{Mock_results.fig}
 \vspace{3mm}
\end {figure*}

Then, we fit each of the simulated observed-frame RC-realizations with an exponential disk model as done with the real data, and we also apply the same selection (i.e. Criterion 2) to include galaxies for stacking.  After selection, we average the noise-free theoretical RCs at a given [$R_{\rm{e}}$,$m_{\rm{d}}$] according to their normalization, accounting for the typical FOV limitations in our data.  Finally, we measure the outer slope on this average for a given [$R_{\rm{e}}$,$m_{\rm{d}}$] within similar radii as done for the stacked rotation curve discussed in Section\ \ref{shape.sec}.

Figure C1 plots a few observed-frame mock-RC examples, showing four realizations for different adopted $m_{\rm{d}}$, with an indication of {whether or not a RC was rejected}.  As apparent from Figure C1, the outer slope of a simulated RC {is strongly correlated} with $m_{\rm{d}}$.  In cases of strong dark matter dominance ($m_{\rm{d}} = 0.01$), a {strong} outer flattening is still present in the outer observed mock-RC (in cases where $R_{\rm{obs}}$ is sufficiently large), which can be reasonably well recovered by the exponential disk model shown in red.  {Our simulations also show that the primary reason for a galaxy to be rejected is that its rotation curve cannot be traced to sufficiently large radii (i.e. $R_{\rm{obs}}$ is too small).}\\
To approach {Question 1 above} we first examine the fraction of observed-frame mock-RCs we selected for stacking as a function of $m_{\rm{d}}$, displayed in the left panel of  Figure C2.  Overall, the fraction of selected galaxies over the entire range of $m_{\rm{d}}$ is $\sim 45 - 60 \%$ and shows a mild trend with $m_{\rm{d}}$ as indicated by the dashed line.  The fraction of galaxies selected for stacking from the \kd + SINS/zC-SINF AO samples presented in Section\ \ref{stacking_sample.sec}, {is} in rough agreement with the numbers found here.  Since we found that the variation of $R_{\rm{obs}}$ in the mock-RC realizations is the {dominant} factor for determining whether a galaxy gets rejected or not, we conclude that the main effect setting the selected fraction of galaxies for stacking out of the \kd + SINS/zC-SINF AO samples is the $S/N$ level (setting $R_{\rm{obs}}$) and its variation among the data.  However, as revealed by our analysis here, a mild trend of the 
selection 
fraction and $m_{\rm{d}}$ remains, implying 
that dark matter poor galaxies will be preferentially selected.  To test the effect arising from this trend on our results, we undertake the following exercise:\\
First we consider a large sample of galaxies with an underlying log-normal distribution of $m_{\rm{d}}$ with 0.4 dex scatter and apply the selection function shown in the left panel of Figure C2 to the distribution of $m_{\rm{d}}$.  Then we determine the resulting average $m_{\rm{d}}$ by again randomly drawing from the distribution. We find that the resulting bias in $m_{\rm{d}}$ is small, such that the selection function in Figure C2 changes the average $m_{\rm{d}}$ by $\sim ~4 \%$ at most over the entire range of $m_{\rm{d}}$ values tested.  {Although this test is simplistic}, it indicates that even when rejecting a large number of {galaxies for stacking}, this does not bias the average $m_{\rm{d}}$ of galaxies in the selected sample {strongly}.\\
Averaging over the entire range of $m_{\rm{d}}$, we find a positive correlation between the fraction of selected galaxies and $R_{\rm{e}}$, such that larger galaxies are more preferentially selected (ranging from $\sim 51\%$ for $R_{\rm{e}}= 2$kpc to $\sim 58\%$ for $R_{\rm{e}}= 5$kpc).  This trend results from the built-in correlation between $R_{\rm{obs}}$ and $R_{\rm{e}}$ in our mock analysis here, and is furthermore in qualitative agreement with the bias towards larger galaxies for our stacking sample compared to the underlying \kd + SINS/zC-SINF AO samples.

Next, we turn to answering {Question 2 above} by examining the outer slopes measured on our simulated mock rotation curves. The right panel of Figure C2 displays the recovered outer slopes determined from the selected, normalized and averaged mock-RCs in the full grid of [$R_{\rm{e}}$,$m_{\rm{d}}$], shown as circles. In addition, the outer slopes determined from the noise-free rotation curves with a given [$R_{\rm{e}}$,$m_{\rm{d}}$] are shown as dashed lines (i.e. which have \textit{not} been perturbed by noise, normalized, selected, and averaged).  All slopes are shown in units of {normalized} coordinates $V/V_{\rm{max}}$ and $R/R_{\rm{turn}}$, {as for} the analysis of our results in Section \ref{structure.sec}.  As expected, the theoretical outer slope of the mock-RCs is a strong function of $m_{\rm{d}}$, with a mild dependence on $R_{\rm{e}}$.  

The method of {selecting}, normalizing and averaging our mock-RCs can reproduce {the} outer slopes within the range in $R_{\rm{e}}$ probed, most importantly even in the 
regime of rising rotation curves.  
The differences between the theoretical and recovered values are small, considering the uncertainty in the normalized slope of our stack ($\sim 0.1 [V/V_{\rm{max}}$,$R/R_{\rm{turn}}]$).  Since the dashed lines {represent} all 25 realizations for a given [$R_{\rm{e}}$,$m_{\rm{d}}$], the difference between the circles and the dashed lines also encompasses the bias in slope due to selection of galaxies discussed above, further substantiating that this bias is small.\\
To investigate possible mass dependencies, we repeat our entire analysis changing the baryonic mass of the disk by $\pm$ 0.5 dex and find that this does not influence our conclusions made here. \\

Although the analysis presented here is solely based on 1D-profiles and thus provides only a simplified view {of} our full stacking methodology, it {demonstrates} the robustness of our stacking technique and sample selection.  {With regard to Question 1}, our analysis {reveals} that the selection of our stacking sample out of the \kd + SINS/zC-SINF AO samples is unlikely to cause strong biases in $m_{\rm{d}}$.  We find that the selection of galaxies for our stacking is mostly driven by the radius out to which individual RCs can be constrained, i.e. {set by the $S/N$} ratio for a given galaxy.\\
{With regard to} Question 2, {our simulations show that our stacking technique is able to recover a wide range of outer slopes.}  Although the model we use to normalize individual RCs intrinsically falls off in rotation velocity after the turn-over, {we are still able to recover positive outer slopes and the large dark matter fractions they imply.}


\end {document}